\documentclass[english]{article}
\usepackage[T1]{fontenc}
\usepackage[latin9]{inputenc}
\usepackage{geometry}
\usepackage{float}
\usepackage{amsmath}
\usepackage{amssymb}
\usepackage{graphicx}
\usepackage{natbib}
\setcitestyle{authoryear,open={(},close={)}} 
\usepackage[colorinlistoftodos]{todonotes}
\usepackage{enumitem}

\makeatletter
\usepackage{chngcntr}
\counterwithout{figure}{section}

\makeatother

\usepackage{babel}

\newcommand{\RomanNumeralCaps}[1]
    {\MakeUppercase{\romannumeral #1}}
    
\usepackage{setspace}

\begin{document}

\begin{center} \LARGE\textbf{Structural features of microvascular networks trigger\\ blood-flow oscillations}\\[5mm] \normalsize Y. Ben-Ami$^1$, G. W. Atkinson$^1$, J. M. Pitt-Francis$^2$, P. K. Maini$^1$, H. M. Byrne$^1$\\[3mm] \end{center} \small 1 Wolfson Centre for Mathematical Biology, Mathematical Institute, University of Oxford, Oxford, United Kingdom\\[2mm] 2 Department of Computer Science, University of Oxford, Oxford, United Kingdom \begin{center} \footnotesize October 13, 2021  \end{center} \normalsize

\begin{abstract}
We analyse mathematical models in order to understand how microstructural features of vascular networks may
affect blood-flow dynamics, and to identify particular
characteristics that promote the onset of self-sustained oscillations.
By focusing on a simple three-node motif, we predict that network ``redundancy'',
in the form of a redundant vessel connecting two main flow-branches,
together with differences in haemodynamic resistance in the branches,
can promote the emergence of oscillatory dynamics. We use existing mathematical descriptions for blood rheology and haematocrit splitting at vessel
branch-points to construct our flow model; we combine numerical simulations
and stability analysis to study the dynamics of the three-node network
and its relation to the system's multiple
steady-state solutions. While, for the case of equal inlet-pressure conditions, a ``trivial'' equilibrium solution
with no flow in the redundant vessel always exists, we find that it
is not stable when other, stable, steady-state attractors exist.
In turn, these ``nontrivial'' steady-state solutions may undergo
a Hopf bifurcation into an oscillatory state. We use the branch diameter
ratio, together with the inlet haematocrit rate, to construct a two-parameter stability
diagram that delineates regimes in which such oscillatory dynamics
exist. We show that flow oscillations in this network geometry
are only possible when the branch diameters are sufficiently different
to allow for a sufficiently large flow in the redundant vessel, which acts as
the driving force of the oscillations. These microstructural properties,
which were found to promote oscillatory dynamics, could be used to
explore sources of flow instability in biological microvascular networks.
\end{abstract}

\section{Introduction \label{sec: introduction}}

Sustained oscillations in the microcirculation have been
known to occur for some time. They have been observed \emph{in
vivo} \citep{Kim96} and \emph{\textit{in vitro} }\citep{Fro12},
and studied via mathematical modelling (for example, \cite{Kia94,CL00,Ged10}). Such
theoretical studies have confirmed the existence of oscillatory solutions that
are self-induced, i.e., they emerge in the absence of external forcing or the imposition of oscillatory boundary conditions. In this work we seek to identify the microstructural features of vascular networks which promote the oscillatory dynamics. Understanding such relationships is important
in order to predict the behaviour of large-scale vascular networks
and the tissue oxygenation they provide.
While several theoretical studies have demonstrated that network geometry can affect the emergence of blood flow oscillations (see, for example, \cite{Ged07, DP14a, KS15}), to our knowledge no explicit mechanism has been proposed for how specific structural features of microcapillary networks, together with the inherent properties of blood flow, act to generate oscillatory dynamics. Therefore, in this work, we consider a simple three-node network motif as a model case study to investigate how geometrical features of the network promote oscillatory instability of steady blood flow. 
Our findings could be used in future work to identify motifs in larger networks that act as sources of instability and trigger oscillatory dynamics.

When blood flows in microcapillaries its viscosity is governed by the concentration of red blood cells (RBCs), such that the hydraulic resistance depends nonlinearly on the haematocrit level, a phenomenon known as the F\aa{}hr\ae{}us-Lindqvist
effect \citep{FL31}. On the other hand, the splitting of haematocrit at a vessel branch point depends (nonlinearly) on the partitioning of blood flow between daughter vessels, a phenomenon known as ``plasma skimming'' \citep{Kro21}. Together, these two effects result in coupled nonlinear relations between haematocrit concentrations and flow rates in the different vessels of the network. Such intrinsic nonlinearities have been shown to drive the emergence of multiple equilibria and oscillatory dynamics \citep{KS15}.

The complex rheology of microcapillary blood flow and its strong dependence
on the haematocrit and vessel diameter
was originally studied by \cite{FL31}.
Later on, \cite{Pea94} derived a widely-used
mathematical model to quantify these effects. While the \cite{Pea94} viscosity model is a fundamental element of almost
every study of blood flow in the microcirculation, a variety
of models for plasma skimming have been used. The haematocrit
splitting rules vary from simple, single-parameter equations \citep{KJ82,FCC85}
to complicated models based on experimental measurements \citep{Pri89}
and discrete-RBC simulations \citep{Ber20}. 

The functional form of the haematocrit splitting models has been
shown to significantly affect the emergence of self-sustained oscillations;
specifically, a dominant factor is the rate at which the haematocrit flux into a specific daughter branch increases as the total flow-rate into that branch increases. For example, \cite{DP11,DP14a,DP14b} showed that regular networks, such as honeycomb
or tree networks, are only prone to oscillations when physically unrealistic
splitting rules, with very large haematocrit-flux gradients, are used. Nevertheless,
other research \citep{KS15,Gar10} has shown that when the more biologically-sound
haematocrit splitting model of \cite{Pri89} is used,
multiple equilibria and oscillations may occur if some ``redundancy''
is introduced into the system; this was achieved by adding a vessel
to connect two main flow branches. The term ``redundancy'' is used
here because, in certain conditions, the additional connecting vessel can support a ``zero-flow'' solution where it transports neither plasma nor haematocrit.
These findings suggest that self-induced oscillations are more likely
to occur in vascular networks with irregular topological structures,
which are, in fact, also characteristic of tumour vasculature \citep{Jain05}.

Cancer cells influence and respond to local environmental conditions; this leads to rapid and localised angiogenesis,
and the formation of networks whose morphologies differ dramatically
from those of healthy tissues \citep{Jain05}. It is hypothesised
that the abnormal structure of tumour vasculature leads to spatio-temporal
variations in blood flow and haematocrit distribution, which manifest
at the macroscopic level as cycling hypoxia \citep{CTF16,Gil18}. This phenomenon is characterized by periodic episodes of oxygen deprivation, followed by periods of reoxygenation, in localised tumour regions. Tumour
cells exposed to such fluctuating hypoxia dynamics experience a selective
advantage for malignant growth \citep{Hoc96} and resistance to chemo-
and radio-therapy \citep{HB04,Gra53,Hor12}. In spite of the obvious
impact on tumour behaviour, the mechanisms and structural
irregularities that contribute to oscillatory tumour
blood flow remain unclear. We postulate that the existence of many
redundant vessels in tumour networks, combined with the intrinsic
nonlinearities of microscale blood flow, can play a significant role in such tumour blood-flow fluctuations.

As a first step towards better understanding the microscale
mechanisms that lead to unsteady flows and cycling hypoxia
in tumours, we revisit the simplest model for network redundancy --
a three-node network in which two main flow branches are connected by
a redundant vessel. Motivated by irregular tumour networks, where regulatory angiogenic mechanisms may be disrupted, and a range
of vascular diameters may prevail, we consider different branch diameters
in our model. We will show that such differences can have a significant affect on the existence of oscillatory
dynamics. We combine numerical simulations and stability analysis
to study the blood flow dynamics of the three-node network. By varying the branch diameter ratio (representing structural effects) and the inlet haematocrit level (representing the effect of local conditions), we construct
a two-parameter stability diagram that delineates regimes in which
multiple equilibria and oscillatory solutions exist. Using the haematocrit
splitting model of \cite{Pri89}, we demonstrate
how features of the model, particularly its nonsmoothness, affect the emergence of oscillatory instability.

In Sec.~\ref{sec:Blood-Flow in capillary network} we describe the model for blood flow in a three-node network and introduce the method we use to simulate its time evolution. In Sec.~\ref{sec:triangle_network} we use dynamic simulations and linear stability analysis to characterise the steady and dynamic behaviours of the flow in the network. We summarise our conclusions in Sec. \ref{sec:Conclusion}. Technical details relating to the analysis
carried out in Sec.~\ref{sec:triangle_network} are presented in
the Appendix.

\section{Model for unsteady blood flow in a three-node capillary network\label{sec:Blood-Flow in capillary network}}

We study the unsteady flow of blood through a series of cylindrical capillaries that form a three-node network. The dependent variables are the nodal pressures, vessel flow-rates, and haematocrit distributions.
We start in Sec.~\ref{sec:single_vessel} by formulating the coupled equations describing the flow dynamics in a single vessel as a function of its inlet and outlet conditions. Then, in Sec.~\ref{network_model}, we formulate the three-node network model, describing how the blood flow and haematocrit in the different vessels are related at vascular junction points (internal nodes). A computational algorithm for simulating the time evolution of the flow in the network is given in Sec.~\ref{sec:dynamic}.

\subsection{Blood-flow dynamics in a single vessel \label{sec:single_vessel}} 

We start by describing the flow in a single vessel. Since we focus on blood flow in microcapillaries, we neglect inertial effects and consider viscous flow in a cylindrical vessel whose length, $L$, is much larger than its diameter, $D$.
Under the assumption of radial symmetry, the flow is assumed to follow Poiseuille's law which, after averaging over the vessel length, $L$, takes the form:

\begin{equation}
\triangle p(t) = \overline{R}(t)Q(t),\label{def_R}
\end{equation}

where $Q(t)$ is the total volumetric
flow-rate and $\triangle p(t) = p(x_{0},t)-p(x_{L},t)$ is the pressure drop along the vessel ($x_{0}$ and $x_{L}$ denote the vessel entrance and end points,
respectively; for straight vessels: $L=x_{L}-x_{0}$). Additionally,

\[
\overline{R}(t)=\frac{128}{\pi}\frac{\overline{\mu}(t) L}{D^{4}},
\]

represents the vessel-averaged haemodynamic resistance, and  $\overline{\mu}(t)$ denotes the vessel-averaged viscosity which is given by

\begin{equation}
\overline{\mu}(t)=\frac{1}{L}\int_{x_{0}}^{x_{L}}\mu\left(H(x,t),D\right)\mathrm{d}x.\label{eq:avg_viscosity}
\end{equation}


%

Following \cite{Pea94}, we assume that the apparent blood viscosity, $\mu$, depends on the vessel diameter, $D$ (given in units of microns) and the discharge haematocrit, $H=H(x,t)$, represents the ratio of RBC flux to total flow rate, as follows:

\begin{equation}
\mu(H,D)=\eta\beta\left[1+\beta\left(\eta_{45}-1\right)\frac{\left(1-H\right)^{c}-1}{\left(1-0.45\right)^{c}-1}\right].
\label{eq:viscosity}
\end{equation}

In Eq.~(\ref{eq:viscosity}),

\[
\beta=\left(\frac{D}{D-1.1}\right)^{2},\;\;\;\eta_{45}=6e^{-0.085D}+3.2-2.44e^{-0.06D^{0.645}},
\]

\[
C=\left(0.8+e^{-0.075D}\right)\left(\frac{1}{f}-1\right)+\frac{1}{f},
\]

and

\[
f=1+10\left(\frac{D}{10}\right)^{12}.
\]

In order to calculate the average viscosity in Eq.~(\ref{eq:avg_viscosity}), the spatio-temporal distribution of haematocrit has to be evaluated. When considering haematocrit transport we assume the case of a plug flow,
i.e., the radially-averaged velocity of RBCs is equal to the radially-averaged plasma velocity (we neglect the F\aa{}hr\ae{}us effect \citep{Far29}). While this assumption
is made to simplify the analysis, we note that \cite{KS15} have shown that including the F\aa{}hr\ae{}us effect does not significantly impact the system dynamics. Under these assumptions the haematocrit
distribution within each vessel is governed by the following one-dimensional advection equation:

\begin{equation}
\frac{\partial H}{\partial t}+U(t)\frac{\partial H}{\partial x}=0,\label{eq:H_mass_balance}
\end{equation}

where the radially-averaged blood velocity $U(t)$ is given by
\[
U(t)=\frac{4Q(t)}{\pi D^{2}}.
\]
By integrating Eq.~(\ref{eq:H_mass_balance}) along the length of the vessel, it is straightforward to show that the vessel-averaged haematocrit,

\[
\overline{H}(t)=\frac{1}{L}\int_{x_{0}}^{x_{L}}H(x,t)\mathrm{d}x,
\]

is such that

\begin{equation}
\frac{\mathrm{d}\overline{H}}{\mathrm{d}t}=\frac{U(t)}{L}\left[H\left(x_{0},t\right)-H(x_{L},t)\right].\label{eq:dH_avg_dt}
\end{equation}

Similarly, averaging Eq.~(\ref{eq:viscosity}) and differentiating with respect to time, we deduce that the average viscosity, $\overline{\mu}(t)$, evolves as follows

\begin{equation}
\frac{\mathrm{d}\overline{\mu}}{\mathrm{d}t}=\frac{U(t)}{L}\frac{\eta\beta^{2}\left(\mu_{45}-1\right)}{\left(1-0.45\right)^{c}-1}\left[\left(1-H\left(x_{0},t\right)\right)^{C}-\left(1-H(x_{L},t)\right)^{C}\right].\label{eq:d_mu_avg_dt}
\end{equation}

Equations (\ref{eq:dH_avg_dt}) and (\ref{eq:d_mu_avg_dt})
can be solved if $H\left(x_{0},t\right)$ and $H(x_{L},t)$
are known. 

It is straightforward to deduce from Eq.~(\ref{eq:H_mass_balance}) that $H(x,t)$ is constant along the characteristic curves

\begin{equation}
\frac{\mathrm{d}x}{\mathrm{d}t}=U(t).\label{eq:charecteristcs}
\end{equation}
Integrating Eq.~(\ref{eq:charecteristcs}) with respect to $t$, we have that

\begin{equation}
x_L-x_0-\int_{t-\tau(t)}^{t}U(s)\mathrm{d}s=0,
\label{eq:int_U_tau}
\end{equation}

where $\tau(t)$ is the time taken for haematocrit to propagate along a vessel of length $L=x_L-x_0$.
The inlet haematocrit $H(x_{0},t)$ depends on the haematocrit in the parent vessel(s) (see Sec.~\ref{network_model}). Since the haematocrit is constant along the characteristic curves, if the inlet haematocrit in a specific vessel is known, then the corresponding outlet haematocrit is given by

\begin{equation}
H(x_{L},t)=H\left(x_{0},t-\tau(t)\right),\label{eq:H_L}
\end{equation}

where $\tau(t)$ is defined implicitly by Eq.~(\ref{eq:int_U_tau}). In cases for which $\int_{0}^{t}U(s)\mathrm{d}s<L$ (the initial inlet haematocrit, $H(x_0,t=0)$, has yet to propagate along the length of the vessel), we assign initial conditions for the outlet haematocrit, $H(x_L,t)=H(x_L,t=0)$.

We note that if the velocity changes sign
during a simulation, additional complexities can arise. For example, in some cases the haematocrit may not have reached the vessel end-point before it starts to propagate backwards. These complexities will not be discussed here, for the sake of brevity, and because no such sign changes occurred for any of the simulations of the three-node network reported in the current work.

\subsection{Three-node network model\label{network_model}}

\begin{figure}[H]
\begin{centering}
\includegraphics[scale=0.5]{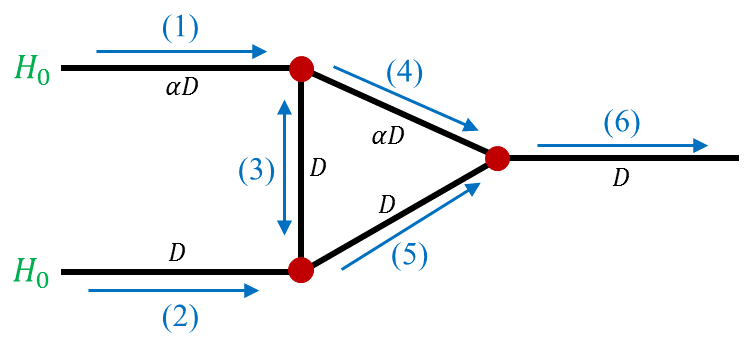}
\par\end{centering}
\caption{Schematic of the three-node network. The blue arrows indicate the possible
flow directions in each vessel. All vessels have the same length $L$, while
the vessel diameters ($D$ for vessels 2, 3, 5 and 6; 
$\alpha D$ for vessels 1 and 4) are indicated in black. }
\label{Fig._triangle}
\end{figure}

We consider unsteady blood flow within a three-node microcapillary network in order to derive a simple model to investigate how structural features of the network may invoke multistability and oscillatory dynamics. The three-node network consists of six vessels (numbered 1 to 6), all of length $L$, and has two inlets and a single outlet (see Fig.~\ref{Fig._triangle}). Motivated by the irregular structure of tumour vasculature, we focus on the effect of asymmetry in the diameters of the two main flow branches. We assume that the diameters of vessels 2, 3, 5 and 6 are identical and denote by $D$ their diameters. We assume further that the diameters of vessels 1 and 4 are identical and denote by $\alpha$ the ratio of their diameters to those of vessels 2, 3, 5 and 6. As we show below, this difference in vessel diameters plays an important role in the emergence of oscillatory dynamics in the three-node network.

Blood flow dynamics are greatly affected by the local haematocrit level due to its affect on blood rheology. While macroscale haematocrit levels are generally uniform in vascular networks ($H \approx 0.45$ in humans), microscale haematocrit levels can be extremely heterogeneous due to plasma skimming effects \citep{Pri92}. In order to study the combined effect of geometrical features and local haematocrit conditions, we assign a constant haematocrit discharge $H_{0}$ at both inlets. In this study, the inlet haematocrit, $H_{0}$, and the diameter ratio, $\alpha$, serve as the two key parameters governing the system dynamics.

We impose constant and equal pressure differences between both inlets and the outlet nodes: a discussion of alternative boundary conditions is included in Sec~\ref{subsec:trivial}.

\begin{figure}[H]
\begin{centering}
\includegraphics[scale=0.5]{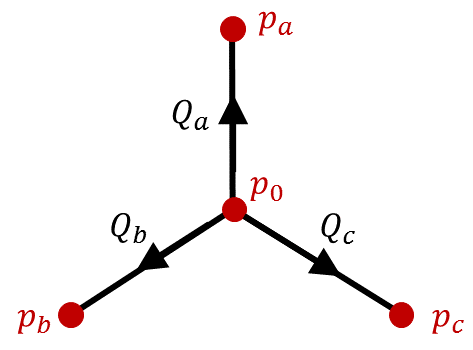}
\par\end{centering}
\caption{Schematic of a generic vascular junction point. Arrows indicate the direction assigned with a positive flux in a particular vessel.}

\label{Fig_junction}
\end{figure}

In Sec.~\ref{sec:single_vessel} we considered flow in a single vessel, assuming that the boundary conditions (pressure difference and inlet haematocrit) for this vessel could be determined from the flow in other vessels. We now formulate a system of algebraic equations for the nodal pressures based on mass conservation applied at internal nodes.
Each internal node (marked with red circles in Fig.~\ref{Fig._triangle}) represents a junction between three vessel segments (see Fig.~\ref{Fig_junction}).
Accordingly, application of conservation of mass at an internal node yields


\begin{equation}
Q_a+Q_b+Q_c=0,\label{eq:sum_Q}
\end{equation}

where the indices $a$, $b$, and $c$ refer to fluxes from the central node, where the pressure is $p_0$, towards connected nodes with pressures $p_a$, $p_b$, and $p_c$, respectively, as illustrated in Fig.~\ref{Fig_junction}. This notation is generic, i.e., it does not relate to a specific junction; later we will use the notation $Q_i$ ($i=1,2..,6$) to denote fluxes in specific vessels within the three-node network.
Combining Eqs.~(\ref{def_R}) and (\ref{eq:sum_Q}) we arrive at the following equation for the nodal pressure at a generic junction

\begin{equation}
p_{0}=\frac{K_{a}p_{a}+K_{b}p_{b}+K_{c}p_{c}}{K_{a}+K_{b}+K_{c}},\label{eq:equation_for_p}
\end{equation}


where

\[
K_{j}=\frac{\pi}{128}\frac{D_{j}^{4}}{L \overline{\mu}_{j}},\;\;j=a,\ b,\ c
\]

represents the vascular conductivity ($1/R_j$ in Eq.~(\ref{def_R})) as a function of the specific diameter, $D_{j}$, and average viscosity, $\overline{\mu}_{j}$, in the vessel connecting node 0 and the nodes marked with $j=a,b,c$. If we apply Eq.~(\ref{eq:equation_for_p}) at all internal nodes of the network, and prescribe the pressures of the inlet and outlet nodes, then we obtain a system of algebraic equations for the nodal pressures.
In order to close these equations, knowledge of the average viscosity within all network vessels, $\overline{\mu}_{i}$ ($i=1,2,...,6$), is required. In order to determine the viscosity, we must also calculate the corresponding discharge haematocrit values, $H_i(x,t)$.

In Sec.~\ref{sec:single_vessel} we studied haematocrit propagation through a single vessel. For a network, we must relate the haematocrit at the entrance to that vessel and the haematocrit at the end of the parent vessel (or vessels) that supply it with haematocrit. For example, consider the node connecting vessels 2, 3, and 5 in Fig.~\ref{Fig._triangle} (denoted as node 2-3-5 from now on). If the flows through vessels 2 and 3 converge into vessel 5, then the haematocrit at the entrance to vessel 5 is determined by applying a mass
balance to the RBCs,

\begin{equation}
H_{5}(x_{0},t)=\frac{H_{2}(x_{L},t)Q_{2}(t)+H_{3}(x_{L},t)Q_{3}(t)}{Q_5(t)}.\label{eq:junction_mass_balance}
\end{equation}

Here, the vessel edges $x_{0}=0$ and $x_{L}=L$ refer to points
in the coordinate system for a specific vessel ordered so that the pressure at $x_{0}$ is higher than that at $x_{L}$. Alternatively, if the flux in vessel 3 changes direction such that the flows in vessels
3 and 5 diverge from vessel 2, then a haematocrit splitting
rule is applied,

\begin{equation}
H_{3}(x_{0},t)=\frac{Q_2 H_{2}}{Q_{3}}F_{3}|_{(x_{L},t)}\ \ \ \text{and}\ \ \ H_{5}(x_{0},t)=\frac{Q_2 H_{2}}{Q_{5}}F_{5}|_{(x_{L},t)}, \label{eq:HS_junction}
\end{equation}

where the functions $F_{3}$ and $F_{5}=1-F_{3}$ (due to haematocrit conservation) define the splitting rules for vessels 3 and 5, respectively. In the simplest nonlinear models,
e.g., the model of \cite{KJ82}, the splitting
rule depends solely on the flux ratio between the daughter and
parent branches. More generally, however, the splitting rule may also depend on the branch diameters and haematocrit. In
this work we use the model of \cite{Pri89}. For the three-node network under consideration, haematocrit splitting only occurs at either node 2-3-5 or node 1-3-4, and we denote the splitting function at node 2-3-5 as $F_3=F(\psi)$, where $\psi=Q_3/Q_2$, while the splitting function at node 1-3-4 is $F_3=F^*(\psi^*)$, where $\psi^*=Q_3/Q_1$. For illustrative purposes, here we define the splitting function at node 2-3-5:

\begin{equation}
F(\psi)=\begin{cases}
0, & \psi<\psi_{0}\\
\frac{e^{A}\left(\psi-\psi_{0}\right)^{B}}{e^{A}\left(\psi-\psi_{0}\right)^{B}+\left(1-\psi-\psi_{0}\right)^{B}} & \psi_{0}\leq\psi\leq1-\psi_{0}\\
1, & \psi>1-\psi_{0}
\end{cases}\label{eq:Pries_model}
\end{equation}

where

\[
A=-\frac{6.96}{D_{2}}\ln\left(\frac{D_{3}}{D_{5}}\right),\ \ \ B=1+6.98\frac{1-H_{0}}{D_{2}},\;\;\;\text{and}\;\;\;\psi_{0}=\frac{0.4}{D_{2}}.
\]

In Eq.~(\ref{eq:Pries_model}) we have used the fact that the haematocrit in vessel 2 is equal to the inlet haematocrit, $H_0$.
The splitting function in node 1-3-4, $F^*(\psi^*)$, can be readily obtained by replacing $D_2$ by $D_1$ and $D_5$ by $D_4$ in the splitting function's coefficients $A$, $B$, and $\psi_0$, following Eq.~(\ref{eq:Pries_model}).

It is important to note that at bifurcations, the conserved quantities are the total flow, $\sum_{i=1}^{3}Q_{i}=0$, and the haematocrit flow, $\sum_{i=1}^{3}Q_{i}H_{i}=0$, but not the haematocrit concentration. Due to the nonlinear form of the splitting rule, the proportion of haematocrit flux bifurcating into the favoured daughter branch may be larger than the proportion of total flux bifurcating into this branch (`plasma skimming'), leading to increased haematocrit concentration in the favoured daughter branch relative to the parent branch. The non-linearity of the splitting rule facilitates non-uniform distribution of haematocrit between the different vessels.

With all the components for modelling blood flow within a microcapillary network defined, we now explain how we construct model solutions for the nodal pressures, vessel flow-rates, and average haematocrit.

With the haematocrit distributions known at an initial time, the nodal pressures can be found using Eq.~(\ref{eq:equation_for_p}), and the flux (and average velocity) in each vessel can be determined using Eq.~(\ref{def_R}). Then, the inlet haematocrit for each vessel can be deduced from the outlet haematocrit of its parent vessel(s) (\ref{eq:HS_junction}) or (\ref{eq:junction_mass_balance}). Evaluating the haematocrit propagation time, $\tau$, using Eq.~(\ref{eq:int_U_tau}), and assigning to Eq.~(\ref{eq:H_L}), the right-hand-sides of Eqs.~(\ref{eq:dH_avg_dt}) and~(\ref{eq:d_mu_avg_dt}) are determined, such that the time evolution of the average haematocrit $\overline{H}$ and viscosity $\overline{\mu}$ can be calculated. The numerical algorithm for implementing the outlined model is given below.

\subsection{Dynamic simulation algorithm \label{sec:dynamic}}

The following algorithm was used to generate numerical solutions for the time-dependent flow
and haematocrit propagation in the capillary network:
\begin{enumerate}
\item The initial inlet haematocrit $H_{i}(x_{0},0)$, outlet haematocrit
$H_{i}(x_{L},0)$, and average haematocrit $\overline{H}_{i}(0)$, are prescribed for each vessel ($i=1,2,...,6$), such that they are
equal at a specific vessel. The corresponding initial values of the average viscosity, $\overline{\mu}_{i}(0)$, are then calculated in each vessel. A constant haematocrit value, $H_{0}$, is prescribed in the inlet vessels. In such vessels: $H(x_{0},t)=H(x_{L},t)=\overline{H}(t)=H_{0}$.

Constant pressures, $p_{\mathrm{in}}=\triangle P$ and $p_{\mathrm{out}}=0$, are prescribed at the inlet and outlet nodes, respectively, such that a constant overall pressure-difference, $\triangle P$, is imposed between both inlet nodes and the outlet node.

\item We adopt the methodology employed by \cite{DP11}
to solve Eq.~(\ref{eq:equation_for_p}) for the internal nodal pressures at
each time step using a Gauss-Seidel iterative method. Iterations
are continued until the relative change in value of each nodal pressure on subsequent iterations is less than a prescribed tolerance of $10^{-20}$.

\item The fluxes $Q_i$ and average velocities $U_i$ are calculated for all vessels ($i=1,2,...,6$)
using Eq. (\ref{def_R}).

\item At each timestep ($n=1,2,...,N$, $t^n=n\triangle t$; $\triangle t$ denotes the magnitude of the time increment), and in each vessel ($i=1,2,...,6$), the haematocrit propagation time, $\tau_{i}^{n}$, is calculated by numerically integrating Eq.~(\ref{eq:int_U_tau}). The integral in Eq.~(\ref{eq:int_U_tau})
is evaluated for $t-\tau$ set equal to each discrete time in
the interval $\left[0,t^{n}\right]$. The approximated value of $t-\tau$
is then determined as the time when the left-hand side of Eq. (\ref{eq:int_U_tau})
changes its sign. When the flow in the vessel does not change its
direction, the search for the value of $t-\tau$ can be restricted
to the interval $\left[(t-\tau)_{i}^{n-1},t^n\right]$, where $(t-\tau)_{i}^{n-1}$
is the retarded time found in the previous timestep. This feature is crucial in order to obtain reasonable run
times of the simulation code.

\item The haematocrit at each vessel end point, $H_i(x_L,t)$, is evaluated using Eq. (\ref{eq:H_L}).

\item Depending on the flow directions in the vessels connected to each internal node, either a haematocrit flux balance (\ref{eq:junction_mass_balance})
or a splitting rule (\ref{eq:HS_junction}) is used to update
the inlet haematocrit value, $H_i(x_0,t)$, in the daughter branch(es).

\item The average haematocrit and viscosity values in each vessel are numerically
advanced in time by applying Euler quadrature to Eqs. (\ref{eq:dH_avg_dt})
and (\ref{eq:d_mu_avg_dt}).

\item Steps 2-6 are repeated until $t=t^N$.
\end{enumerate}

The above algorithm was implemented in MATLAB. The code is available at the following GitHub repository: https://github.com/yaronbenami/blood\_flow

\section{Characterising the steady and dynamic behaviours of a three-node
network\label{sec:triangle_network}}

In order to streamline the analysis of the three-node network considered in Fig.~\ref{Fig._triangle}, we nondimensionalise the governing equations. Fluxes are normalised by the steady-state flow rate in
vessel 2 so that

\[
Q_i=\frac{\widetilde{Q}_i}{\widetilde{Q}_{2}^{(0)}},
\]

where the superscript $(0)$ denotes a steady-state value and, hereafter,
tildes denote dimensional quantities. When pressure boundary conditions are prescribed, $\widetilde{Q}_{2}^{(0)}$ changes as the system parameters $\alpha$ and $H_0$ vary. However, this scaling was chosen because it renders the dimensionless formulation of the haematocrit splitting rule less cumbersome.

Average vessel resistances, as defined by Eq.~(\ref{def_R}),
are scaled by the constant resistance in vessel 2,

\[
\overline{R}_i=\frac{\pi\widetilde{D}^{4}}{128\widetilde{\mu}\left(H_{0},D\right)\widetilde{L}}\overline{\widetilde{R}}_i = \frac{\overline{\mu}\left(H_{i},\alpha_i D\right)}{\alpha_i^4\mu\left(H_{0},D\right)},
\]

where $D=\widetilde{D}/1 \mu m$ as appropriate for using the viscosity function of \cite{Pea94} (Eq.~(\ref{eq:viscosity})), $\alpha_i=\alpha$ for $i=1,4$ and $\alpha_i=1$ otherwise (see Fig. \ref{Fig._triangle}).

The spatial coordinate is scaled by $\widetilde{L}$, such
that for each vessel $x\in[0,1]$; time is nondimensionalised by the time for haematocrit to propagate through vessel 2 when the flow is steady

\[
t=\frac{\widetilde{U}_{2}^{(0)}}{\widetilde{L}}\widetilde{t}.
\]

Under these scalings, the dimensionless parameters governing the system behaviour are

\[
\alpha,\;\;H_0,\;\;\text{and}\;\;D,
\]

and the parameter values used in this work are such that 
$\alpha \in[0.25,2.25]$,  $H_0 \in[0,1]$, and $D=20$.

\subsection{Steady-state solutions \label{subsec:steady_state_bifurcation}}

For the three-node network, if the same pressure is imposed
at both inlets, then a steady solution with no flow in vessel 3 always
exists. In this case, the haematocrit in all vessels is equal to $H_{0}$, except for vessel 3, which has no haematocrit. It is straightforward
to show that the dimensionless flux in the upper branch (scaled by $\widetilde{Q}_{2}^{(0)}$) is given by

\begin{equation}
Q_{1}^{(0)}=Q_{4}^{(0)}=\frac{1}{R_{1}^{(0)}}=\frac{\alpha^{4}\mu\left(H_{0},D\right)}{\mu\left(H_{0},\alpha D\right)}.\label{eq:Q_1_trivial}
\end{equation}

Henceforth, we refer to this steady solution as the ``trivial solution'', and
vessel 3 as the ``redundant vessel'', because there is always a steady state for which it transports neither haematocrit nor plasma. With $Q_3^{(0)} = H_3^{(0)} = 0$ for the trivial solution, the pressure drops along vessels 1 and 4 (and 2 and 5) are equal (i.e., $\triangle p_{1}^{(0)}=\triangle p_{4}^{(0)}$
and $\triangle p_{2}^{(0)}=\triangle p_{5}^{(0)}$) for all values of the parameters
$\alpha$ and $H_{0}$. If we impose equal inlet-pressure
conditions, then the pressure drop along the redundant vessel is $\triangle p_{3}^{(0)}=0$, which is consistent with the definition of the trivial solution as a state for which $Q_{3}^{(0)}=0$.

In previous work, \cite{Gar10} showed that the three-node network admits multiple steady-state solutions. They also showed that the network possesses three solutions if the inlet haematocrit, $H_0$, exceeds a threshold value. However, they did not characterise the stability of the steady-state solutions.
Therefore, in this section, we characterise the multiple steady-state solutions
of the network in order to subsequently study their stability characteristics (in Sec.~\ref{subsec:trivial}. \emph{et seq.}) We start by formulating
the equations that define the two non-trivial steady-state solutions:

\begin{itemize}
\item Case \RomanNumeralCaps{1} -- flux
flows from the bottom to the top branch (node 2-3-5 to node 1-3-4). 
\item Case \RomanNumeralCaps{2} -- flux
flows from the top to the bottom branch (node 1-3-4 to node 2-3-5). 
\end{itemize}

These solutions differ in the node at which the haematocrit splitting rule is imposed: for Case \RomanNumeralCaps{1}, the haematocrit splitting rule is imposed at node 2-3-5, while for Case \RomanNumeralCaps{2} it is imposed at node 1-3-4.

At a steady state, the haematocrit in each vessel is independent of spatial position $x$ (we set $\partial/\partial t=0$ in Eq. (\ref{eq:H_mass_balance})),
and, thus, the resistance is also independent of $x$. Further, the pressure difference in vessel $i$ depends on the flux via

\begin{equation}
\triangle p_{i}^{(0)}=R_{i}^{(0)}Q_{i}^{(0)},\label{eq:delta_p_q_steady}
\end{equation}

where, from Eq. (\ref{def_R}),

\begin{equation}
R_{i}^{(0)}=\frac{\mu(H_{i}^{(0)},\alpha_{i}D)}{\mu\left(H_{0},D\right)\alpha_{i}^{4}}\label{R_steady}.
\end{equation}

If we consider the steady-state pressure drop along the loop formed by the three internal nodes, then we have that

\begin{equation}
\triangle p_{3}^{(0)}+\triangle p_{4}^{(0)}-\triangle p_{5}^{(0)}=0.\label{eq:dp_triangle-1}
\end{equation}

For Case \RomanNumeralCaps{1} (i.e., flux flows from the bottom to the top branch), substituting (\ref{eq:delta_p_q_steady}) and (\ref{R_steady}) into (\ref{eq:dp_triangle-1}) yields

\begin{equation}
\begin{gathered}\frac{\mu(H_{4}^{(0)},\alpha D)}{\alpha^{4}}Q_{1}^{(0)}+\left(\mu(H_{3}^{(0)},D)+\frac{\mu(H_{4}^{(0)},\alpha D)}{\alpha^{4}}+\mu(H_{5}^{(0)},D)\right)Q_{3}^{(0)}-\mu(H_{5}^{(0)},D)=0.\end{gathered}
\label{eq:triangle_steady_Q3_negative}
\end{equation}

In Eq.~(\ref{eq:triangle_steady_Q3_negative}) we have also exploited the mass balance Eq. (\ref{eq:sum_Q}) at nodes 1-3-4 and 2-3-5, which yields $Q_{4}^{(0)}=Q_{1}^{(0)}+Q_{3}^{(0)}$
and $Q_{5}^{(0)}=1-Q_{3}^{(0)}$, respectively.
Additionally, imposition of equal inlet pressures yields

\begin{equation}
\triangle p_{1}^{(0)}-\triangle p_{3}^{(0)}-\triangle p_{2}^{(0)}=0.\label{eq:dp_inlets}
\end{equation}

Substituting (\ref{eq:delta_p_q_steady}) into (\ref{eq:dp_inlets}), together with the prescription of inlet haematocrit $H_0$, we find that

\begin{equation}
\frac{\mu(H_{0},\alpha D)}{\alpha^{4}}Q_{1}^{(0)}-\mu(H_{3}^{(0)},D)Q_{3}^{(0)}-\mu(H_{0},D)=0.\label{eq:equal_inlet_steady_Q3_negative}
\end{equation}

Equations (\ref{eq:triangle_steady_Q3_negative}) and (\ref{eq:equal_inlet_steady_Q3_negative})
should be supplemented by expressions for the haematocrit in vessels
3, 4, and 5. The haematocrit splitting rule (\ref{eq:Pries_model})
at node 2-3-5 can be written as

\begin{equation}
H_{3}^{(0)}=\frac{F(Q_{3}^{(0)})}{Q_{3}^{(0)}}H_{0}.\label{eq:H_3_steady_Q3_negative}
\end{equation}
Equation (\ref{eq:H_3_steady_Q3_negative}), together with haematocrit mass
balance (\ref{eq:junction_mass_balance}), yields the following expression for the haematocrit in vessel 5,

\begin{equation}
H_{5}^{(0)}=H_{0}\left(\frac{1-F(Q_{3}^{(0)})}{1-Q_{3}^{(0)}}\right),
\label{H_3_H_5_steady_Q3_negative}
\end{equation}

while the haematocrit mass balance at node 1-3-4 yields the following expression for the haematocrit in vessel 4,

\begin{equation}
\begin{gathered}H_{4}^{(0)}=H_{0}\left(\frac{Q_{1}^{(0)}+F(Q_{3}^{(0)})}{Q_{1}^{(0)}+Q_{3}^{(0)}}\right).\end{gathered}
\label{eq:H_4_steady_Q3_negative}
\end{equation}

Equations (\ref{eq:triangle_steady_Q3_negative})-(\ref{eq:H_4_steady_Q3_negative}) define the nontrivial steady-state solution for Case \RomanNumeralCaps{1}, where blood flows from the bottom to the top branch.
For Case \RomanNumeralCaps{2}, where blood flows from the top to the bottom branch, a similar analysis leads to the following system of equations:

\begin{equation}
\frac{\mu(H_{4}^{(0)},\alpha D)}{\alpha^{4}}Q_{1}^{(0)}-\left(\mu(H_{3}^{(0)},D)+\frac{\mu(H_{4}^{(0)},\alpha D)}{\alpha^{4}}+\mu(H_{5}^{(0)},D)\right)Q_{3}^{(0)}-\mu(H_{5}^{(0)},D)=0,\label{eq:trinagle_steady_Q3_positive}
\end{equation}

\begin{equation}
\frac{\mu(H_{0},\alpha D)}{\alpha^{4}}Q_{1}^{(0)}+\mu(H_{3}^{(0)},D)Q_{3}^{(0)}-\mu(H_{0},D)=0,
\label{eq:equal_pressure_steady_Q3_positive}
\end{equation}

\begin{equation}
\begin{gathered}
H_{3}^{(0)}=H_{0} \frac {Q_{1}^{(0)}}{Q_{3}^{(0)}} F^{*}(Q_{3}^{(0)}/Q_{1}^{(0)}) ,\ \ \ H_{4}^{(0)}=H_{0}\left(\frac{Q_{1}^{(0)}-F^{*}(Q_{3}^{(0)}/Q_{1}^{(0)})}{Q_{1}^{(0)}-Q_{3}^{(0)}}\right),\\
\text{and}\ \ H_{5}^{(0)}=H_{0}\left(\frac{1+Q_{1}^{(0)}F^{*}(Q_{3}^{(0)}/Q_{1}^{(0)})}{1+Q_{3}^{(0)}}\right).\label{eq:H_steady_Q3_positive}
\end{gathered}
\end{equation}

Here, we have exploited conservation of mass at nodes 1-3-4 and 2-3-5, which now means that $Q_{4}^{(0)}=Q_{1}^{(0)}-Q_{3}^{(0)}$
and $Q_{5}^{(0)}=1+Q_{3}^{(0)}$.  We note that for both Cases \RomanNumeralCaps{1} and \RomanNumeralCaps{2} we assume $Q_{3}^{(0)}>0$, thus the change of flux direction manifests via sign changes between Eqs. (\ref{eq:triangle_steady_Q3_negative}), (\ref{eq:equal_inlet_steady_Q3_negative}) and (\ref{eq:trinagle_steady_Q3_positive}), (\ref{eq:equal_pressure_steady_Q3_positive}), respectively.

\begin{figure}[H]
\begin{centering}
\includegraphics[scale=0.5]{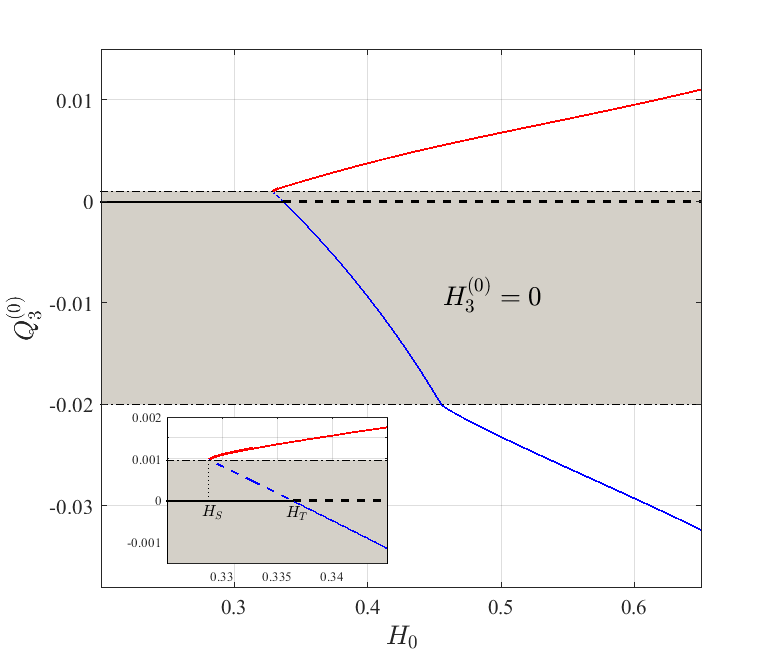}
\par\end{centering}
\caption{Steady state solutions for $Q_{3}^{(0)}$ as a function of the inlet
haematocrit, $H_{0}$, when $\alpha=0.45.$ Each solution branch is
illustrated with a different line-colour, with the black line representing
the trivial solution. Solid and dashed lines correspond to stable and unstable solutions, respectively. The region of Hopf
instability analysed in Secs.~\ref{subsec:Oscillatory-state}-\ref{subsec:Oscillations-_smooth}
is not indicated in this figure. The grey area bounded by the dash-dotted
lines is the region in which the steady-state solutions have no haematocrit
in vessel 3. The inset shows an enlarged image of the two bifurcations: (i) the saddle-node bifurcation, $H_S$, from which the
red and blue solutions originate; (ii) the transcritical bifurcation, $H_T$,
at which the blue and black (trivial) solutions exchange stability.}

\label{Fig_steady_state_solutions}
\end{figure}

Equations (\ref{eq:triangle_steady_Q3_negative})-(\ref{eq:H_4_steady_Q3_negative})
or (\ref{eq:trinagle_steady_Q3_positive})-(\ref{eq:H_steady_Q3_positive})
can be solved numerically to obtain the steady-state solutions in
terms of the system parameters $\alpha$ and $H_{0}$. Naturally,
both sets of equations reduce to the trivial solution when
$Q_{3}^{(0)}=0$. As the trivial solution exists for all parameter values, a bifurcation should occur when one of the nontrivial solutions for $Q_{3}^{(0)}$ approaches $Q_{3}^{(0)}=0$. The strategy we use to identify such bifurcation points is to first find nontrivial steady-state solutions to Eqs. (\ref{eq:triangle_steady_Q3_negative})-(\ref{eq:H_4_steady_Q3_negative}) and (\ref{eq:trinagle_steady_Q3_positive})-(\ref{eq:H_steady_Q3_positive})
for sufficiently large values of $H_{0}$ -- one for each direction
of flow in the redundant vessel.
Then, we use numerical continuation to track these solution-branches
as $H_{0}$ decreases. In Fig.~\ref{Fig_steady_state_solutions} we present the multiple
steady-state solutions found using this numerical tracking technique
for $\alpha=0.45$. For all calculations,
we assumed a nominal, dimensionless diameter of $D=20$ (which corresponds to a dimensional diameter of 20 $\mu m$). As mentioned above, for both Cases \RomanNumeralCaps{1} and \RomanNumeralCaps{2},
$Q_{3}^{(0)}$ is considered positive. However, to illustrate the two flow directions in Fig. \ref{Fig_steady_state_solutions}, we denote $Q_{3}^{(0)}<0$ as the solution associated with Case \RomanNumeralCaps{1} and $Q_{3}^{(0)}>0$ as the solution associated with Case \RomanNumeralCaps{2}.

For fixed values of $\alpha$, tracking the nontrivial solutions as
$H_{0}$ decreases eventually leads to one of the following scenarios:

\begin{enumerate}[label=(\roman*)]

\item If the flux in the redundant vessel is directed
towards the higher resistance vessel (i.e., $Q_{3}^{(0)}<0$; solid blue line in Fig.~\ref{Fig_steady_state_solutions}), then the nontrivial and trivial solutions (blue and black curves in Fig.~\ref{Fig_steady_state_solutions},
respectively) meet at a transcritical bifurcation, denoted $H_{T}$ ($H_{T}=0.3365$ in Fig.~\ref{Fig_steady_state_solutions}).

\item If the flux in the redundant vessel is directed
towards the lower resistance vessel (i.e., $Q_{3}^{(0)}>0$; red curve in Fig.~\ref{Fig_steady_state_solutions}),
then the nontrivial solution branch ceases to exist at a saddle-node (fold) bifurcation at which $H_{0}=H_S<H_T$ ($H_{S}=0.3288$ in Fig.~\ref{Fig_steady_state_solutions}).

\end{enumerate}

The thin dash-dotted black lines in Fig.~\ref{Fig_steady_state_solutions}
denote the critical fluxes, $|Q_{3}^{(0)}|=0.4/D$ (bottom
line) and $|Q_{3}^{(0)}/Q_{1}^{(0)}|=0.4/(\alpha D)$ (top line) at which haematocrit propagation through the redundant vessel is initiated. These lines indicate where the flux ratio between the redundant vessel and its parent vessel (vessel 2 for the bottom line and vessel 1 for the top line) attains the critical value $\psi_{0}$ in the haematocrit splitting model (\ref{eq:Pries_model}) for the two flow configurations.
Only where the ratio of the fluxes in vessels 3 and 2 (bottom line) or 3 and 1 (top line) exceeds these threshold values, will haematocrit enter vessel 3. Thus, $H_{3}^{(0)}=0$ in the grey region bounded by these two lines. Following \cite{KS15}, we term these critical values ``skimming thresholds''.

The function used in Eq. (\ref{eq:Pries_model}) imposes
nonsmoothness of the splitting rule at the skimming threshold. The emergence of a fold bifurcation at one of the skimming thresholds
(dash-dotted black lines in Fig.~\ref{Fig_steady_state_solutions}),
might lead to the conjecture that the nonsmoothness of the haematocrit
splitting function is responsible for the emergence of the fold bifurcation.
However, a separate analysis, using a smooth splitting rule (see Sec.~\ref{subsec:Oscillations-_smooth}),
yields a qualitatively similar bifurcation diagram which, for the
sake of brevity, is not presented here. We conclude that the
bifurcation structure shown in Fig.~\ref{Fig_steady_state_solutions} is due to the network configuration, rather than the specific splitting rule.

\subsection{Stability of the trivial solution\label{subsec:trivial}}

\begin{figure}[H]
\includegraphics[scale=0.37]{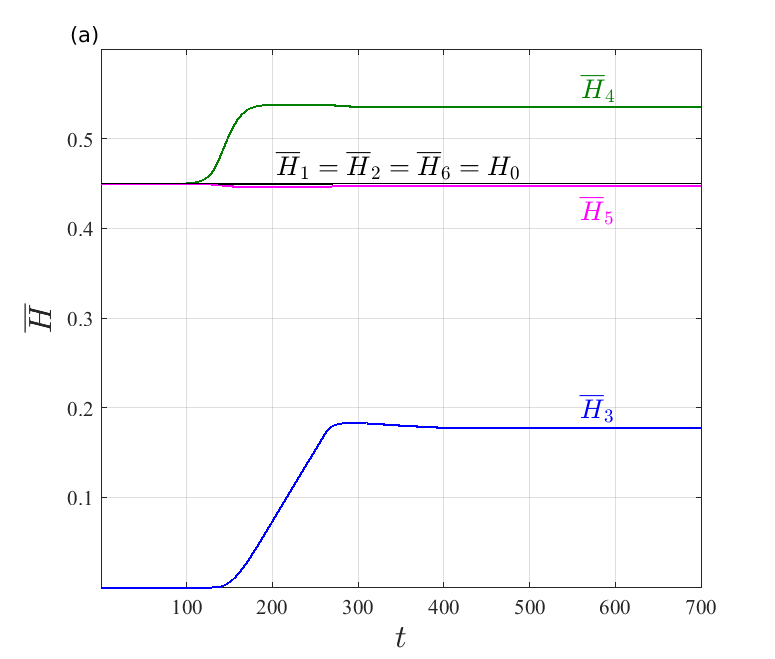}\hfill{}\includegraphics[scale=0.37]{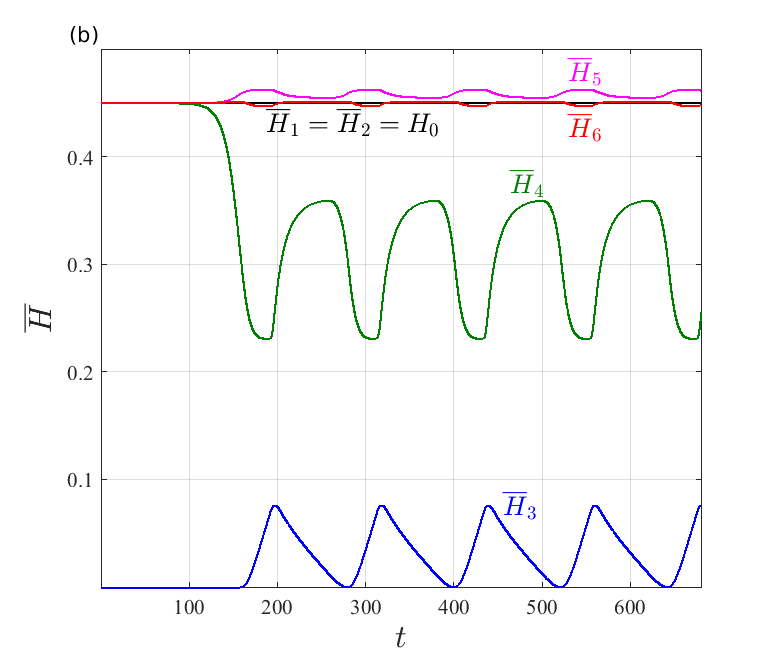}

\vfill{}

\includegraphics[scale=0.37]{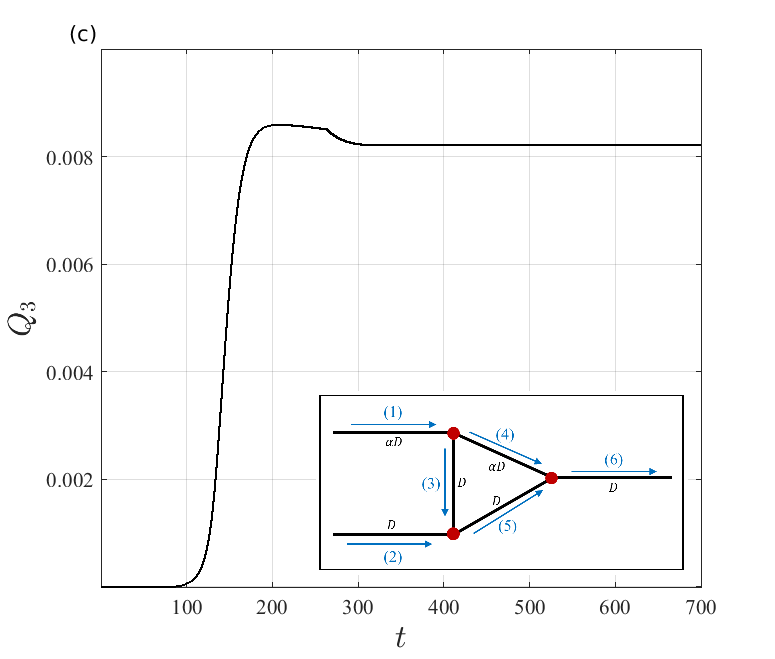}\hfill{}\includegraphics[scale=0.37]{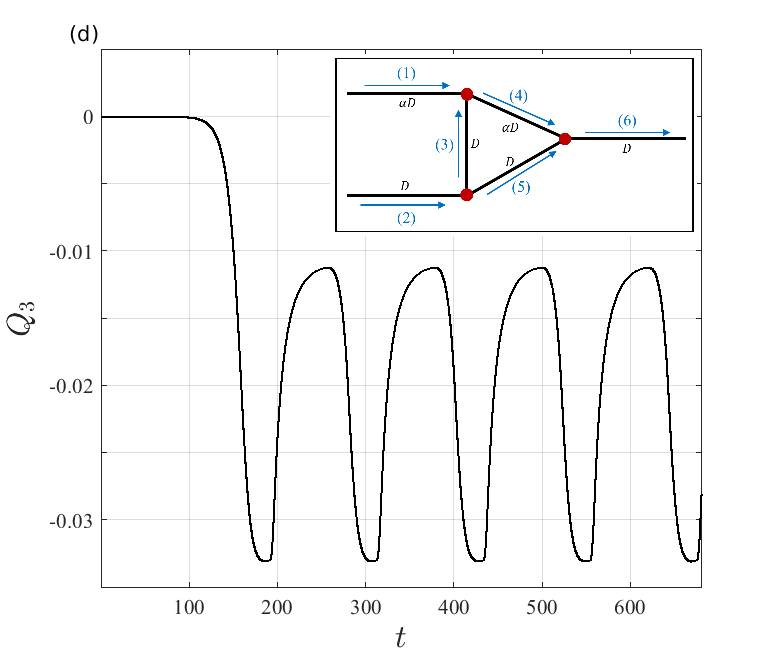}

\caption{Simulation results 
showing that the trivial steady-state is unstable. When perturbed, the system evolves either to a different steady state solution (a),(c) or to oscillatory dynamics (b),(d), the long time dynamics depending on the sign of the haematocrit perturbation. The time evolution of $\overline{H}$, the average
haematocrit (a),(b) and $Q_3$, the fluid flux in the redundant vessel (c),(d), are presented
for different perturbations from the trivial solution with $\alpha=0.5$ and $H_{0}=0.45$.
The insets in (c),(d) illustrate the direction of flow in each vessel; the different flow directions in vessel 3 lead
to either a steady state in (a),(c) or a stable limit cycle in (b),(d).}
\label{Fig_dynamic_simulation}
\end{figure}

The bifurcation diagram presented in Sec.~\ref{subsec:steady_state_bifurcation} suggests that the
trivial solution exchanges stability with nontrivial steady-state solutions at a transcritical bifurcation point.
To verify this finding, we conducted dynamic flow simulations, using the trivial solution to initialise the network flow. 
Figure~\ref{Fig_dynamic_simulation} shows how the vessel-averaged haematocrit (Fig.~\ref{Fig_dynamic_simulation}
(a),(b)) and fluid flux in the redundant vessel (Fig.~\ref{Fig_dynamic_simulation}
(c),(d)) change over time when we fix the ratio of the vessel diameters and
inlet haematocrit so that $\alpha=0.5$ and $H_{0}=0.45$. The simulations evolve to either a different (nontrivial) steady state (Fig. \ref{Fig_dynamic_simulation}(a),(c)) or an oscillatory
state (Fig.~\ref{Fig_dynamic_simulation}(b),(d)). 
The two different dynamics were obtained by imposing small perturbations ($\pm10^{-6}$) on the inlet haematocrit
of vessel 2 at the first time step; a positive perturbation resulted in evolution to a nontrivial steady state (Fig.~\ref{Fig_dynamic_simulation}
(a),(c)), while a negative perturbation resulted in the onset of oscillatory dynamics (Fig.~\ref{Fig_dynamic_simulation}
(b),(d)).
The insets
in Fig.~\ref{Fig_dynamic_simulation}(c),(d) illustrate the different flow directions
in vessel 3. A positive
flux in the redundant vessel $Q_3>0$, i.e., blood flows from the higher ($\alpha<1$) to the lower resistance branch, leads to a nontrivial
steady state (corresponding to the red line in Fig. \ref{Fig_steady_state_solutions}),
while $Q_3<0$ may lead to oscillatory dynamics.

In light of these findings, we performed a linear stability analysis
of the trivial steady-state solution in order to characterise its local stability. We made the following ansatz for the flux, average resistance, and haematocrit in the $i$-th vessel:

\begin{equation}
\begin{gathered}Q_{i}(t)=Q_{i}^{(0)}+\epsilon q_{i}\exp(\lambda t)+O(\epsilon^{2})\;\;,\;\;\overline{R}_{i}(t)=R_{i}^{(0)}+\epsilon r_{i}\exp(\lambda t)+O(\epsilon^{2}),\\
\text{and}\;\;\;H_{i}(x,t)=H_{i}^{(0)}+\epsilon h_{i}\exp\left[\lambda\left(t-\frac{\alpha_{i}^{2}}{Q_{i}^{(0)}}x\right)\right]+O(\epsilon^{2}),
\end{gathered}
\label{eq:anzats}
\end{equation}

where $Q_{i}^{(0)}$, $R_{i}^{(0)}$, and $H_{i}^{(0)}$ represent the trivial steady-state solution given by Eq. (\ref{eq:Q_1_trivial}), with $Q^{(0)}_3=H^{(0)}_3=0$ and $H^{(0)}_{i \neq 3}=H_0$. Further, $\epsilon\ll1$ denotes a small perturbation of the trivial state and
the complex parameter $\lambda=\sigma+\mathrm{i}\omega$ represents
the growth-rate, $\sigma$, and oscillation-frequency, $\omega$, of the perturbation, respectively.
The expression for the haematocrit, in which  $h_i$ represents the $O(\epsilon)$ perturbation to the haematocrit at a vessel inlet $x=0$, satisfies Eq. (\ref{eq:H_mass_balance}) at $O(\epsilon)$.

We note here that the above perturbation equations are for Case \RomanNumeralCaps{1} (blood flows from the bottom to the top branch, i.e.,
$Q_3^{(0)}<0$ in Fig. \ref{Fig_dynamic_simulation}). While
the dynamic simulations have shown that the flow direction in the
redundant vessel affects the attractor to which the system evolves,
we will now show that it does not affect the stability of the trivial solution (the effect of the flow direction in the redundant
vessel will be clarified in Sec.~\ref{subsec:Oscillatory-state},
where the stability of the nontrivial steady-state solutions will
be examined).

To obtain the haematocrit perturbation in vessel 3, we state the haematocrit splitting rule at node 2-3-5:

\begin{equation}
H_{3}(0,t)=\frac{F\left(Q_{3}\right)}{Q_{3}}H_{0}.\label{eq:H_3}
\end{equation}

Substituting from (\ref{eq:anzats}) in (\ref{eq:H_3}), with $Q_3^{(0)}=0$, and equating terms of $O(\epsilon)$, we deduce that

\begin{equation}
h_{3}=0.\label{eq:h_3_trivial}
\end{equation}

This is because when we use the model of \cite{Pri89}, $F(Q_{3}^{(0)}=0)=0$
and $\partial^{n}F/\partial\psi^{n}|_{Q_{3}^{(0)}=0}=0$ for all $n$ (see Eq. (\ref{eq:Pries_model})). We note further that the trivial steady-state is only possible for haematocrit splitting models with $F(Q_{3}^{(0)}=0)$ and $\partial F/\partial\psi|_{Q_{3}^{(0)}=0}=0$.
Additionally, the linear stability of the trivial solution depends
on the value of $\partial^{2}F/\partial\psi^{2}|_{Q_{3}^{(0)}=0}$,
which is identically zero in Eq.~(\ref{eq:Pries_model}). These features
will be important when we discuss the effect of smoothing the splitting
function in Sec.~\ref{subsec:Oscillations-_smooth}.

Applying the haematocrit mass balance, Eq.~(\ref{eq:junction_mass_balance}), at nodes 2-3-5 and 1-3-4, we find that

\begin{equation}
h_{5}=H_{0}q_{3}\;\;\;\text{and}\;\;\;h_{4}=-R_{1}^{(0)}H_{0}q_{3}.\label{eq:h_4_5_trivial}
\end{equation}

In Eq.~(\ref{eq:h_4_5_trivial}) we have used Eq.~(\ref{eq:Q_1_trivial}) from which we have that $Q_{4}^{(0)}=Q_{1}^{(0)}=1/R_{1}^{(0)}$.
Since no haematocrit splitting occurs when we perturb about
the trivial steady-state ($h_3=0$; see Eq. (\ref{eq:h_3_trivial})), Eqs.~(\ref{eq:h_3_trivial})-(\ref{eq:h_4_5_trivial})
are applicable for negative and positive values of $Q_{3}$. This explains why the stability of the trivial state is independent of the flow direction in the redundant vessel.

To complete the haematocrit distribution we apply mass balance
at node 4-5-6 to obtain

\begin{equation}
h_6=\frac{1}{1+R_{1}^{(0)}}\left[h_{4}\exp\left(-R_{1}^{(0)}\alpha^{2}\lambda\right)+R_{1}^{(0)}h_{5}\exp\left(-\lambda\right)\right].
\end{equation}

Having defined the haematocrit perturbations in terms of the flux perturbation $q_3$, we now relate the hydrodynamic-resistance perturbations to the haematocrit perturbations. In the unsteady case the pressure difference is related to the flux via the average resistance (Eq.~(\ref{def_R})), where the nondimensional average resistance reads,

\begin{equation}
\overline{R}_{i}(t)=\frac{\overline{\mu}_{i}(t)}{\mu\left(H_{0},D\right)\alpha_{i}^{4}},
\end{equation}

and

\[
\overline{\mu}_{i}(t)=\int_{0}^{1}\mu\left(H_{i},\alpha_{i}D\right)\mathrm{d}x.
\]
Expanding $\overline{R}_{i}$ as a regular power series in the small parameter $\epsilon \ll 1$, we obtain Eq.~(\ref{R_steady}) at leading order, and at $O(\epsilon)$ we have

\begin{equation}
r_{i}=\frac{Q_{i}^{(0)}}{\alpha_{i}^{6}}\left[1-\exp\left(-\frac{\alpha_{i}^{2}}{Q_{i}^{(0)}}\lambda\right)\right]\frac{1}{\mu\left(H_{0},D\right)}\frac{\mathrm{d}\mu}{\mathrm{d}H}|_{\left(H_{0},\alpha_{i}D\right)}h_{i}.\label{eq:resistance_series}
\end{equation}

Then, we relate the fluxes and resistances in the different vessels by applying Eq.~(\ref{def_R}) to the three constraints on the pressure drops in the network. At $O(\epsilon)$, Eq. (\ref{eq:dp_triangle-1}) reads

\begin{equation}
R_{3}^{(0)}q_{3}+R_{1}^{(0)}q_{4}-q_{5}+\frac{1}{R_{1}^{(0)}}r_{4}-r_{5}=0,\label{eq:dp_triangle_2}
\end{equation}

where we have used the result from Eqs. (\ref{eq:h_3_trivial})
and (\ref{eq:resistance_series}) that $r_{3}=0$.
Since we impose a constant pressure difference $\triangle P$ between the inlet and the outlet nodes, we may assume without loss of generality that the $O(\epsilon)$ perturbation to $\triangle P$ is zero. Then we have that

\begin{equation}
\begin{aligned}
0 = & \triangle p_1^{(\epsilon)} + \triangle p_4^{(\epsilon)} + \triangle p_6^{(\epsilon)},\\
0 = & \triangle p_2^{(\epsilon)} + \triangle p_5^{(\epsilon)} + \triangle p_6^{(\epsilon)},
\end{aligned}
\label{dp_1_and_2}
\end{equation}

where $\triangle p_i^{(\epsilon)}$ ($i=1,2,...,6$) denote $O(\epsilon)$ pressure differences.
Substituting for $\triangle p_i^{(\epsilon)}$ from Eqs.~(\ref{def_R}) and (\ref{eq:anzats}) into Eq.~(\ref{dp_1_and_2}), we deduce that

\begin{equation}
R_{1}^{(0)}(q_{1}+q_{4})+q_{6}+\frac{1}{R_{1}^{(0)}}r_{4}+\left(\frac{1+R_{1}^{(0)}}{R_{1}^{(0)}}\right)r_{6}=0
\label{eq:dp_1}
\end{equation}

and

\begin{equation}
R_{1}^{(0)}q_{1}-q_{2}-R_{3}^{(0)}q_{3}=0,\label{eq:dp_1_eq_dp_2}
\end{equation}

where, since we impose constant inlet haematocrit values ($H_1=H_2=H_0$), we have assumed, without loss of generality, that $h_1=h_2=0$; this leads, via Eq.~(\ref{eq:resistance_series}), to $r_{1}=r_{2}=0$.

Three additional equations are obtained by balancing the flow at each node:

\begin{equation}
\begin{aligned}q_{1}+q_{2}-q_{6}&=0,\\
q_{1}+q_{3}-q_{4}&=0,\\
q_{2}-q_{3}-q_{5}&=0.
\end{aligned}
\label{eq:pressure_BC_last}
\end{equation}

Equations (\ref{eq:h_3_trivial})-(\ref{eq:pressure_BC_last}) form a transcendental eigenvalue problem for $\lambda$. In practice, however, when perturbing about the trivial steady-state solution, $\lambda$ attains only real values. In order to show this, it is helpful to consider the simpler problem of fixed-flux boundary conditions.
In this case, we fix $q_{1}=q_{2}=0$ instead of imposing zero $O(\epsilon)$ pressure drops between the inlet and outlet nodes (Eqs. (\ref{eq:dp_1}) and (\ref{eq:dp_1_eq_dp_2})). Then the eigenvalue problem reduces to a single equation,

\begin{equation}
\begin{aligned}1+R_{3}^{(0)}+R_{1}^{(0)} & = \\
 &\frac{H_{0}}{\mu\left(H_{0},D\right)}\left\{ \frac{1}{\alpha^{4} \lambda_{\alpha}}\left[1-\exp\left(-\lambda_{\alpha}\right)\right]\frac{\mathrm{d}\mu}{\mathrm{d}H}|_{\left(H_{0},\alpha D\right)}+\frac{1}{\lambda}\left[1-\exp\left(-\lambda\right)\right]\frac{\mathrm{d}\mu}{\mathrm{d}H}|_{\left(H_{0},D\right)}\right\},
\end{aligned}
\label{eq:charecteristic_flux_BC}
\end{equation}

where $\lambda_{\alpha}=\alpha^2 R_1^{(0)}\lambda$.
For any choice of $\lambda$, the imaginary parts of the two terms in the right hand side have the form
\[
-C_{\alpha}\omega\left[ 1 - \frac{\sin(\alpha^2 R_1^{(0)}\omega+\theta)}{\sin \theta} \exp(-\alpha^2 R_1^{(0)}\sigma) \right],
\]
and
\[
-C \omega\left[1 - \frac{\sin (\omega+\theta)}{\sin \theta} \exp(-\sigma) \right],
\]
where $\lambda=\sigma+\mathrm{i}\omega$, $\theta = \tan^{-1}(\omega/\sigma)$, and $C$ and $C_{\alpha}$ are positive coefficients. It can be readily shown that these two terms are both positive (when $\omega<0$), both negative (when $\omega>0$), or both zero (when $\omega =0$).
Since the left-hand-side of Eq. (\ref{eq:charecteristic_flux_BC}) has
no imaginary part, it follows that the equation is only satisfied
for real $\lambda$. The same argument can be applied (although
the calculations are more cumbersome) for when pressure boundary conditions are imposed (results not shown). We conclude that the trivial solution cannot undergo a Hopf bifurcation and that transitions to the two attractors shown in Fig.~\ref{Fig_dynamic_simulation} occur when the trivial solution exchanges stability with one
of the nontrivial steady-state solutions, as shown in Fig.~\ref{Fig_steady_state_solutions}.
With $\omega =0$, linear stability analysis of the trivial
solution cannot explain the transition to the oscillatory state shown
in Fig.~\ref{Fig_dynamic_simulation}(b),(d). In Sec.~\ref{subsec:Oscillatory-state}
we demonstrate that the onset of oscillatory dynamics occurs via a Hopf bifurcation from one of the nontrivial steady-state solutions.

In order to find the critical conditions for instability of the trivial steady-state solution, we analyse Eq. (\ref{eq:charecteristic_flux_BC}) in the limit
as $\lambda\rightarrow0$:

\begin{equation}
1+R_{3}^{(0)}+R_{1}^{(0)}-\frac{H_{0}}{\mu\left(H_{0},D\right)}\left[\frac{1}{\alpha^{4}}\frac{\mathrm{d}\mu}{\mathrm{d}H}|_{\left(H_{0},\alpha D\right)}+\frac{\mathrm{d}\mu}{\mathrm{d}H}|_{\left(H_{0},D\right)}\right]=0.\label{eq:charecteristic_flux_BC_lambda_0}
\end{equation}

With $R_1^{(0)}$ and $R_3^{(0)}$ determined by Eq. (\ref{R_steady}), Eq.~(\ref{eq:charecteristic_flux_BC_lambda_0}) defines the critical curve in $(\alpha,H_0)$ parameter space on which the trivial solution loses stability when flux boundary conditions are imposed. By taking the limit as $\lambda\rightarrow0$ of Eqs. (\ref{eq:h_3_trivial})-(\ref{eq:pressure_BC_last}), an additional critical curve is obtained, defining the conditions under which the trivial solution loses stability when pressure boundary conditions are imposed.

The mechanisms driving instability of the trivial steady-state solution can be explained as follows. Suppose, without any loss of generality, that a small flow perturbation, with zero haematocrit, enters redundant 
vessel~(3) from junction~2-3-5 (as mentioned above, since the linear stability of the trivial solution does not depend on the flow direction in the redundant vessel, it suffices to consider this case). Consequently, the haematocrit and resistance in vessel~(4) decrease while those in vessel~(5) increase. This leads, at supercritical conditions, to more flow being redirected towards 
vessel~(4), creating a positive feedback mechanism which destabilises the trivial solution.

At the critical conditions ($\lambda=0$) given by Eq.~(\ref{eq:charecteristic_flux_BC_lambda_0}) (when flux boundary conditions are imposed), the perturbations in the pressure-drop along the three internal nodes (vessels) caused by the increase in the resistance of vessel~(5) and the corresponding decrease in the resistance of vessel~(4) (the terms in brackets in Eq.~(\ref{eq:charecteristic_flux_BC_lambda_0})) are balanced by the pressure drop due to the
flow perturbation in vessels~(3),~(4), and~(5). This condition yields a relationship between a particular value of $\alpha$ and the critical $H_0$ for instability of the trivial solution. For inlet haematocrits that are larger than the critical value, this balance cannot hold due to the increase in the resistance perturbation, resulting in the positive feedback mechanism described above.

\begin{figure}[H]
\begin{centering}
\includegraphics[scale=0.45]{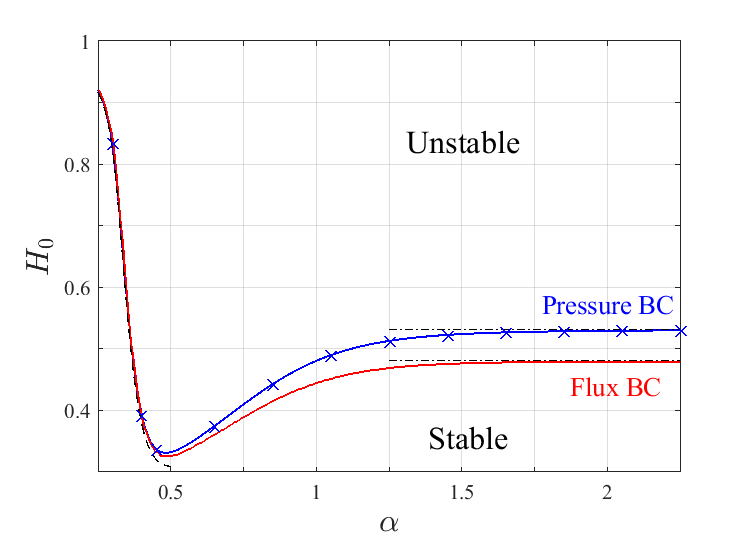}
\par\end{centering}
\caption{Stability diagram of the trivial solution in the $(\alpha,H_{0})$
parameter space. The solid lines mark the critical curves for equal inlet-pressure (Eqs.~(\ref{eq:h_3_trivial})-(\ref{eq:pressure_BC_last}), blue)
and fixed-flux (Eq.~(\ref{eq:charecteristic_flux_BC_lambda_0}), red) boundary conditions. The black dashed
line represents the asymptotic solution for $\alpha\ll1$, while the
black dash-dotted lines mark the asymptotes for $H_{0}$ when
$\alpha\gg1$. The blue crosses indicate the transcritical bifurcation points
of the steady-state solutions ($H_T$) for fixed inlet-pressure conditions
as discussed in Sec.~\ref{subsec:steady_state_bifurcation}.}

\label{Fig_stability_trivial}
\end{figure}

We sketched the critical curves by discretising $\alpha$ in the range $[0.25,2.25]$ and solving numerically for $H_0(\alpha)$ using a continuation scheme, where $\alpha$ is a continuation parameter.
In Fig.~\ref{Fig_stability_trivial} we
plot the critical curves thus obtained for the two types of boundary conditions.
To confirm that the critical curves correspond to the transcritical bifurcation points, $H_{T}$, evaluated in Sec. \ref{subsec:steady_state_bifurcation}, we calculated discrete values of $H_{T}(\alpha)$ by locating the bifurcations in the steady-state solution diagrams for a range of values of $\alpha$. Comparison with the critical curve in Fig.~\ref{Fig_stability_trivial} indicates good agreement, providing independent validation of our stability analysis.

Figure~\ref{Fig_stability_trivial} shows that for a given value of $\alpha$, the critical haematocrit for the
fixed-pressure condition is always larger than the corresponding value for constant-flux conditions, suggesting enhanced instability of the latter. From a mechanistic prospective, we note that under the assumption of Poiseuille flow, the flow rate is proportional to the pressure gradient,

\begin{equation}
Q\propto\frac{\partial p}{\partial x}.
\end{equation}

This means that imposing a flux condition is equivalent to applying
a condition on the derivative of the pressure. Typically, a condition
on a derivative is less restrictive than a condition on the variable itself, leading to reduced parameter range of stability. Our findings are consistent with experimental results reported by \cite{Sto15}. They conducted experiments involving mixtures of two fluids in slightly different network geometries and found that imposing flux boundary conditions produced a larger region of instability than imposing pressure boundary conditions.

We now analyse the stability diagram in the limits of $\alpha\ll1$ ($\alpha\gg1$), representing the cases when the top branch diameter is much smaller (larger) than the diameter of the bottom branch and the redundant vessel.

\subsubsection*{Limit of $\alpha\ll1$}

From Fig.~\ref{Fig_stability_trivial} we note that the critical
stability curves for both types of boundary conditions are similar
when $\alpha\ll1$. In this limit, $R_1^{(0)}=\mu\left(H_{0},\alpha D\right)/\alpha^{4}\mu\left(H_{0},D\right)$ is the dominant steady-state resistance such that

\begin{equation}
R_{3}^{(0)} \sim O(1) \ll R_{1}^{(0)}.
\end{equation}

Additionally, $R_1^{(0)}$ is of a similar magnitude to the resistance perturbations (see Eq.~(\ref{eq:resistance_series})) on both branches,

\begin{equation}
\frac{H_{0}}{\mu\left(H_{0},D\right)}\frac{\mathrm{d}\mu}{\mathrm{d}H}|_{\left(H_{0},D\right)}\sim\frac{H_{0}}{\mu\left(H_{0},D\right)\alpha^{4}}\frac{\mathrm{d}\mu}{\mathrm{d}H}|_{\left(H_{0},\alpha D\right)}\sim R_{1}^{(0)}.
\end{equation}

When flux boundary conditions are imposed, Eq. (\ref{eq:charecteristic_flux_BC_lambda_0})
reduces to

\begin{equation}
\mu\left(H_{0},\alpha D\right)-H_{0}\left[\alpha^{4}\frac{\mathrm{d}\mu}{\mathrm{d}H}|_{\left(H_{0},D\right)}+\frac{\mathrm{d}\mu}{\mathrm{d}H}|_{\left(H_{0},\alpha D\right)}\right]=0,\;\;\;\text{when}\;\;\; \alpha \ll 1. \label{eq:charecteristic_small_alpha}
\end{equation}

When pressure boundary conditions are imposed, analysis of Eqs.~(\ref{eq:h_3_trivial})-(\ref{eq:pressure_BC_last}) in the limit $\alpha \ll 1$ shows that a nonzero solution is only possible if the flux perturbations satisfy the following scaling
\begin{equation}
R_{1}^{(0)}q_{1}\sim q_{2}\sim R_{1}^{(0)}q_{3}.\label{eq:order_of_magnitude_q}
\end{equation}

Assigning scaling (\ref{eq:order_of_magnitude_q}) to equations (\ref{eq:h_3_trivial})-(\ref{eq:pressure_BC_last}) in the limit $\alpha \ll 1$ yields an equation which is identical, at leading order, to Eq.~(\ref{eq:charecteristic_small_alpha}).
The dashed black line in Fig.~\ref{Fig_stability_trivial} corresponds to
the solution of Eq. (\ref{eq:charecteristic_small_alpha}); this line
is indistinguishable from the two solid curves for $\alpha\lesssim0.4$.

\subsubsection*{Limit of $\alpha\gg1$}

It is clear from Fig.~\ref{Fig_stability_trivial} that for both
types of boundary conditions, the critical curves tend to different
constant values of $H_{0}$ when $\alpha\gg1$. In this limit, the steady-state resistance in the branch with diameter $\alpha D$ is small:

\begin{equation}
R_{1}^{(0)}\ll R_{3}^{(0)} \sim O(1),
\label{scale_R_large_alpha}
\end{equation}

and the perturbation to the resistance in this branch is much smaller than the perturbation in the branch with diameter $D$:

\begin{equation}
\frac{1}{\alpha^{4}}\frac{\mathrm{d}\mu}{\mathrm{d}H}|_{\left(H_{0},\alpha D\right)}\ll\frac{\mathrm{d}\mu}{\mathrm{d}H}|_{\left(H_{0},D\right)}.
\label{scale_r_large_alpha}
\end{equation}

Therefore, all the terms which include $\alpha$ in Eq.~(\ref{eq:charecteristic_flux_BC_lambda_0}) for fixed-flux conditions (or Eqs.~(\ref{eq:h_3_trivial})-(\ref{eq:pressure_BC_last}) for fixed-pressure conditions) are negligible
when $\alpha\gg1$, rendering the two sets of equations independent
of $\alpha$, a result which is consistent with the numerical results in Fig.~\ref{Fig_stability_trivial}.
When flux boundary conditions are imposed, Eq. (\ref{eq:charecteristic_flux_BC_lambda_0})
reduces to

\begin{equation}
\frac{H_{0}}{\mu\left(H_{0},D\right)}\frac{\mathrm{d}\mu}{\mathrm{d}H}|_{\left(H_{0},D\right)}=1+R_{3}^{(0)}, \;\;\;\text{when}\;\;\; \alpha \gg 1. \label{eq:large_alpha_flux_BC}
\end{equation}

When pressure boundary conditions are imposed and $\alpha \gg 1$, analysis of Eqs.~(\ref{eq:h_3_trivial})-(\ref{eq:pressure_BC_last}) shows that a nonzero solution is only possible when the flux perturbations are of similar magnitude, that is,

\begin{equation}
q_{1} \sim q_{2} \sim q_{3}.\label{eq:order_of_magnitude_q-1}
\end{equation}

Assigning the scaling in Eq.~(\ref{eq:order_of_magnitude_q-1}) to Eqs.~(\ref{eq:h_3_trivial})-(\ref{eq:pressure_BC_last}) yields

\begin{equation}
\frac{H_{0}}{\mu\left(H_{0},D\right)}\frac{\mathrm{d}\mu}{\mathrm{d}H}|_{\left(H_{0},D\right)}=1+2R_{3}^{(0)},  \;\;\;\text{when}\;\;\; \alpha \gg 1. \label{eq:large_alpha_pressure_BC}
\end{equation}

The black dash-dotted lines in Fig.~\ref{Fig_stability_trivial}
correspond to the asymptotic values of the critical haematocrit, $H_{0}$, when $\alpha\gg1$.
For both types of boundary conditions there is excellent agreement between the asymptotic values of the critical inlet haematocrit for instability and the values given by Eq. (\ref{eq:charecteristic_flux_BC_lambda_0}) or Eqs. (\ref{eq:h_3_trivial})-(\ref{eq:pressure_BC_last}). We note further that, since the left-hand sides of Eqs. (\ref{eq:large_alpha_flux_BC}) and (\ref{eq:large_alpha_pressure_BC})
are monotonically increasing in $H_{0}$, the asymptote for the pressure boundary conditions is larger than that for the flux boundary conditions.

\subsection{Oscillatory solutions \label{subsec:Oscillatory-state}}

Given the oscillatory dynamics predicted by the dynamic simulations
in some cases, it is of interest to identify regions of parameter space in which they
exist. To this end, we performed linear stability analysis of
the nontrivial steady state solutions where, as hinted by Fig.~\ref{Fig_dynamic_simulation},
the flux in the redundant vessel flows from the lower to
the higher resistance branch (this assumption will be justified \textit{a
posteriori}). In order to cover the full range of solutions, we consider separately Cases \RomanNumeralCaps{1} and \RomanNumeralCaps{2} from Sec.~\ref{subsec:steady_state_bifurcation}.
As demonstrated in Sec. \ref{subsec:steady_state_bifurcation},
the key difference between these cases is the node at which the haematocrit splitting rule is imposed.

We use the same ansatz as in Eq. (\ref{eq:anzats}) to perturb the steady-state equations. While the full analysis is presented in Appendices
A and B, important differences between perturbations to the
trivial and nontrivial states are emphasised here. For Case \RomanNumeralCaps{1}, when blood in the redundant vessel flows from the bottom to the top branch (analysis for Case \RomanNumeralCaps{2} is presented in Appendix B), Eqs.~(\ref{eq:triangle_steady_Q3_negative})-(\ref{eq:H_4_steady_Q3_negative}) define the steady-state solutions. Linearising the splitting
function at node 2-3-5 (Eq.~(\ref{eq:H_3})) about its steady state Eq.~(\ref{eq:H_3_steady_Q3_negative}), we deduce that

\begin{equation}
h_{3}=H_{0}\left(\frac{F(Q_{3}^{(0)})}{Q_{3}^{(0)}}-\frac{\partial F}{\partial\psi}|_{Q_{3}^{(0)}}\right)\left[q_{2}-\frac{q_{3}}{Q_{3}^{(0)}}\right].\label{eq:h_3_non_trivial}
\end{equation}

In contrast to the perturbation of the trivial state, here $Q_{3}^{(0)}\neq0$, leading to $h_3 \neq 0$. The haematocrit
mass balance at node 1-3-4 yields the following expression for the perturbation to the haematocrit in vessel 4:

\begin{equation}
\begin{gathered}h_{4}=H_{0}\frac{\left(Q_{3}^{(0)}-F(Q_{3}^{(0)})\right)}{\left(Q_{1}^{(0)}+Q_{3}^{(0)}\right)^{2}}\left[q_{1}-\frac{Q_{1}^{(0)}}{Q_{3}^{(0)}}q_{3}\right]+\frac{Q_{3}^{(0)}}{Q_{1}^{(0)}+Q_{3}^{(0)}}h_{3}\exp\left(-\frac{\lambda}{Q_{3}^{(0)}}\right).\end{gathered}
\label{eq:h_4_non_trivial}
\end{equation}

In Eq.~(\ref{eq:h_4_non_trivial}) we note the exponential term on the right-hand-side, which represents the effect of the time-delay between haematocrit entering vessel
3 and its propagation into vessel 4, induced by the nonzero value
of $Q_{3}^{(0)}$.
This time-delay plays a key role in the emergence of oscillatory dynamics.

\begin{figure}[H]
\begin{centering}
\includegraphics[scale=0.45]{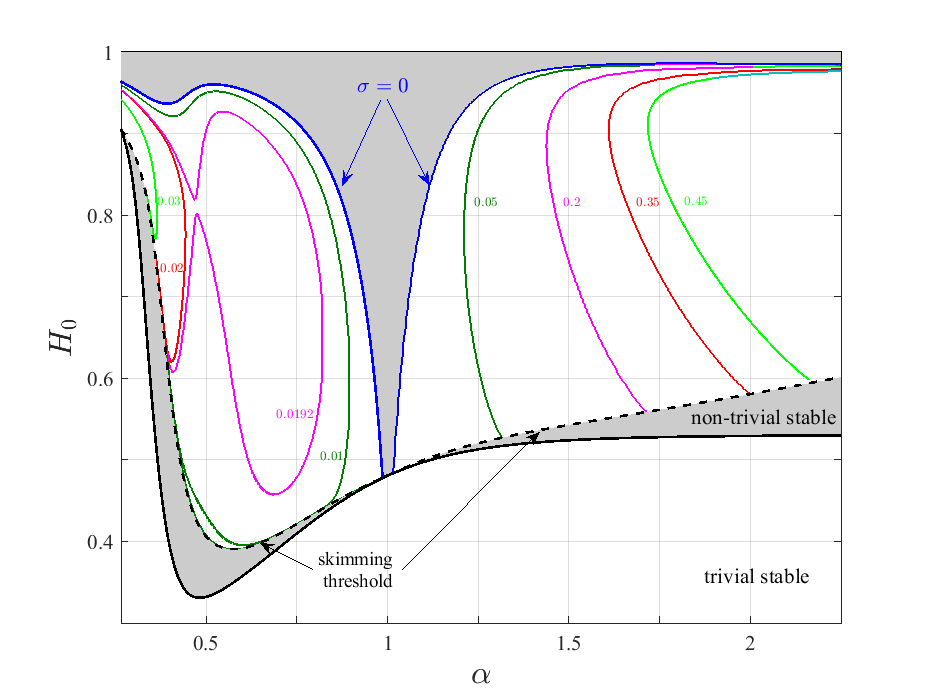}
\par\end{centering}
\caption{Bifurcation diagram indicating regions of $(\alpha,H_{0})$ parameter space in which oscillatory solutions exist when pressure boundary conditions are imposed. The solid black line marks the critical curve for instability of the trivial solution
(and corresponds to the blue curve in Fig.~\ref{Fig_stability_trivial}),
while the dashed black curve corresponds to the skimming threshold of the
nontrivial solutions. The grey areas indicate regions in which both nontrivial
steady states are stable. The coloured curves, all originating from the skimming
threshold, represent iso-growth-rate contours (the values of $\sigma=\mathrm{Re}\{\lambda\}$
are indicated) of the oscillatory solutions; the critical curves on which $\sigma=0$ are marked in blue. The iso-growth-rate contours represent flow
from the bottom to the top branch (Case \RomanNumeralCaps{1}) for $\alpha<1$, and vice versa (Case \RomanNumeralCaps{2}) for $\alpha>1$.}

\label{Fig_oscillatory_nonsmooth}
\end{figure}

The equations governing the perturbations $\mathbf{v}=\{h_3,...,h_6,r_3,...,r_6,q_1,...,q_6\}^{T}$ can be written as a linear system of the form

\begin{equation}
A(\lambda,V^{(0)})\mathbf{v}=0,\label{eq:A_v}
\end{equation}

where the matrix $A(\lambda,V^{(0)})$ depends on the eigenvalue, $\lambda=\sigma+\mathrm{i}\omega$, and the steady-state solution, $V^{(0)}=\{ Q_{1}^{(0)},Q_{3}^{(0)},H_{3}^{(0)},H_{4}^{(0)},H_{5}^{(0)}\} $, which solves Eqs.~(\ref{eq:triangle_steady_Q3_negative})-(\ref{eq:H_4_steady_Q3_negative}) for Case \RomanNumeralCaps{1} (or Eqs.~(\ref{eq:trinagle_steady_Q3_positive})-(\ref{eq:H_steady_Q3_positive}) for Case \RomanNumeralCaps{2}). Full statements of the equations governing the perturbations $\mathbf{v}$ are presented in Appendices A (Case \RomanNumeralCaps{1}) and B (Case \RomanNumeralCaps{2}). 
Equation (\ref{eq:A_v}) constitutes a transcendental eigenvalue problem. In practice, we determine the system's iso-$\sigma$ (iso-growth-rate) contours in the $(\alpha,H_{0})$ parameter space by seeking solutions
for $\{ H_{0},\omega\} $ as a function of $\alpha$ that satisfy

\begin{equation}
\begin{aligned}\mathrm{Re}\left\{ \mathrm{det}\left(A\right)\right\} =0\\
\mathrm{Im}\left\{ \mathrm{det}\left(A\right)\right\} =0
\end{aligned}
\label{eq:determinant}
\end{equation}

for specific values of $\sigma$. To carry out the bifurcation analysis, we used an in-house numerical
continuation scheme to obtain iso-$\sigma$ curves in the $(\alpha,H_0)$ parameter space, with $\alpha$ as the continuation parameter.
We initialised the continuation scheme using solutions estimated from the dynamic simulations presented in Sec.~\ref{sec:dynamic} (for a given combination of values of $\alpha$ and $H_0$, we estimated $\omega$ and $\sigma$ by evaluating the oscillation frequency and growth rate within a short time period after the bifurcation commenced). The resulting
stability diagram is presented in Fig.~\ref{Fig_oscillatory_nonsmooth}, where iso-$\sigma$ curves corresponding to Cases \RomanNumeralCaps{1} ($\alpha<1$) and \RomanNumeralCaps{2} ($\alpha>1$), respectively. The solid black line marks the critical curve for stability of the trivial steady-state solution and the
dashed black line represents the skimming threshold of the redundant
vessel. In the grey region that separates the critical stability curve of the trivial steady-state solution and the skimming threshold, there is no haematocrit in the redundant vessel, although $|Q_{3}^{(0)}|>0$. Remarkably, in this case
the nontrivial steady states are stable; none of the contours with positive growth-rates cross the skimming threshold, indicating that oscillatory instability
can only occur for values of inlet haematocrit $H_0$ above this threshold.
The requirement for haematocrit to be present in the redundant vessel in order to generate oscillatory dynamics can be attributed to the effect of time-delays in
the system: when no haematocrit is present in the redundant vessel
($H_{3}^{(0)}=h_{3}=0$), perturbations in $Q_{3}$ lead to instantaneous
changes in the haematocrit (and, consequently, the resistance) in vessel
4 (or vessel 5 when the flow in vessel 3 is in the opposite direction). Thus, we conjecture that in the absence of time
delays, self-induced oscillations cannot be sustained.

Figure~\ref{Fig_oscillatory_nonsmooth} suggests 
that oscillations only occur
in the presence of multiple equilibria (the oscillatory
regime is always above the critical curve of the trivial solution,
indicating the presence of other nontrivial steady-state solutions).
However, 
the existence of multiple equilibria is not
a necessary condition for the existence of oscillatory states. For
example, \cite{KS15} studied a slightly different network
geometry and identified small regions of parameter space in which oscillatory solutions exist in the presence of only a single steady-state solution. These regions, however, seem to exist only when the diameter of the redundant vessel is very small.

In Fig.~\ref{Fig_oscillatory_nonsmooth}, the iso-$\sigma$ contours do not form closed contours, but they originate from the skimming threshold. We conclude that there is a jump between negative and positive growth rates at the skimming threshold. This discontinuity arises because there is a nonsmooth Hopf bifurcation -- a consequence of the singularity in the system's Jacobian at the skimming threshold, which arises when the exponent $B$ in Eq.~(\ref{eq:Pries_model}) is such that $B<2$. For the parameter
regime used in Fig.~\ref{Fig_oscillatory_nonsmooth}, $B\in[1.14,1.37]$
on the skimming threshold and, hence, we have a nonsmooth
bifurcation (the effect of smoothing the splitting function is discussed
in Sec.~\ref{subsec:Oscillations-_smooth}).
The nonsmooth Hopf bifurcation at the skimming threshold was reported
by \cite{KS15}, who determined the critical conditions
for oscillations in a different network geometry, using
a similar haematocrit splitting model.

Interestingly, Fig.~\ref{Fig_oscillatory_nonsmooth} reveals an additional stable region in of neighborhood of $\alpha=1$ (i.e., where the vessel diameters in the two branches are similar),
suggesting that a critical difference in the flux
between the two branches is needed to trigger oscillatory solutions.
Recall that the two regions in which oscillatory solutions exist in Fig.~\ref{Fig_oscillatory_nonsmooth} correspond to the two flow directions in vessel 3 (oscillations in the bottom-to-top flow configuration are restricted to $\alpha \lesssim 1$, and to $\alpha \gtrsim 1$ when the flow is in the opposite direction).
The fact that the critical curves for the onset of oscillations ($\sigma=0$)
in Fig.~\ref{Fig_oscillatory_nonsmooth} do not cross the line $\alpha=1$
demonstrates that, in this network geometry, oscillatory solutions
can only exist when the flux in the redundant vessel goes from the
lower to the higher resistance branch (see the preliminary
assumptions at the beginning of this section).

The oscillatory dynamics can be understood by considering the two sources of non-linearity: (i) the haematocrit-dependent viscosity of blood (i.e, the
F\aa{}hr\ae{}us-Lindqvist effect), and (ii) the nonlinear splitting
of haematocrit at vessel bifurcations (``plasma skimming''). While
the former induces coupling of the flow and haematocrit concentration,
the latter allows non-uniform haematocrit distributions throughout
the network. Both effects are essential for the feedback mechanisms that
generate self-sustained oscillations: plasma skimming leads to relatively little haematocrit entering the redundant vessel which, in turn, dilutes the haematocrit and, consequently, reduces the resistance to flow (due to the F\aa{}hr\ae{}us-Lindqvist effect) in the smaller-diameter branch,
so that more flow is redirected to the redundant vessel. The time-delayed negative-feedback is also a consequence of the haematocrit-dependent viscosity: the increase of haematocrit in the redundant vessel leads to a delayed increase in the resistance of this flow path. Since no other sources of nonlinearity are included in our model, we believe that it represents a minimal model of how the presence of redundant vessels can promote oscillatory blood flow
when physiologically realistic rules are used to describe the F\aa{}hr\ae{}us-Lindqvist effect and plasma skimming.

\begin{figure}[H]
\includegraphics[scale=0.37]{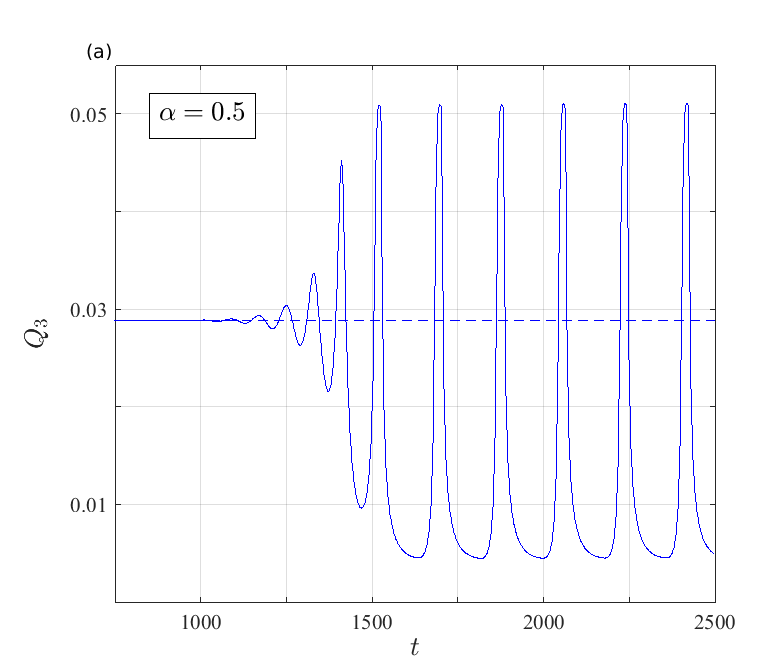}\hfill{}\includegraphics[scale=0.37]{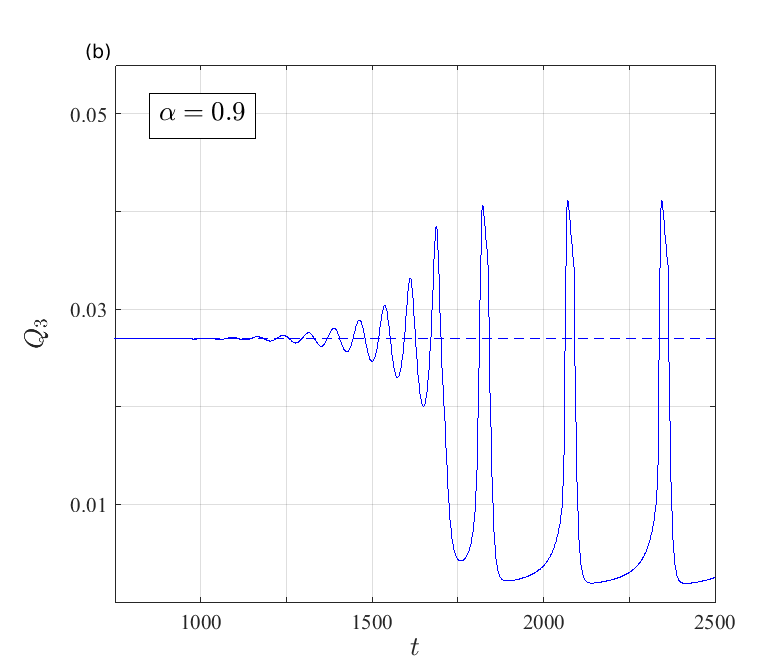}

\vfill{}

\includegraphics[scale=0.37]{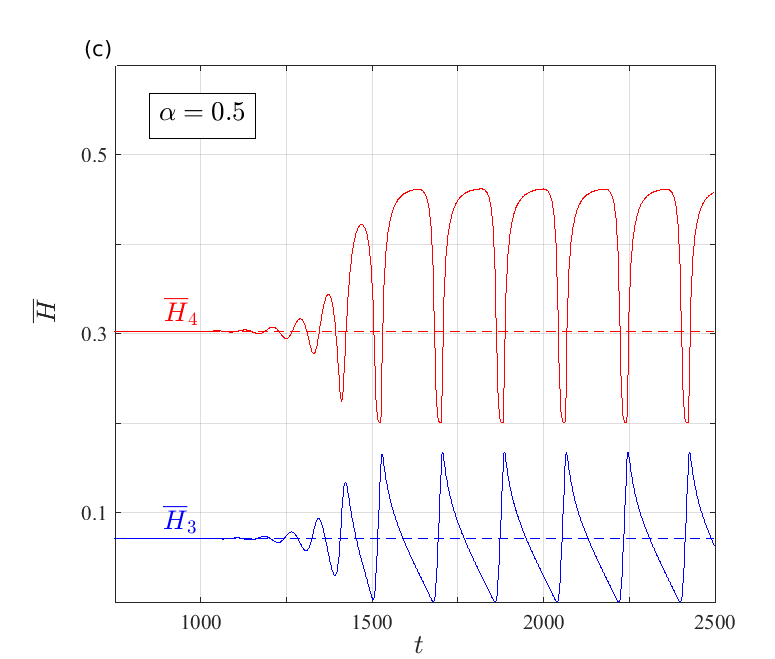}\hfill{}\includegraphics[scale=0.37]{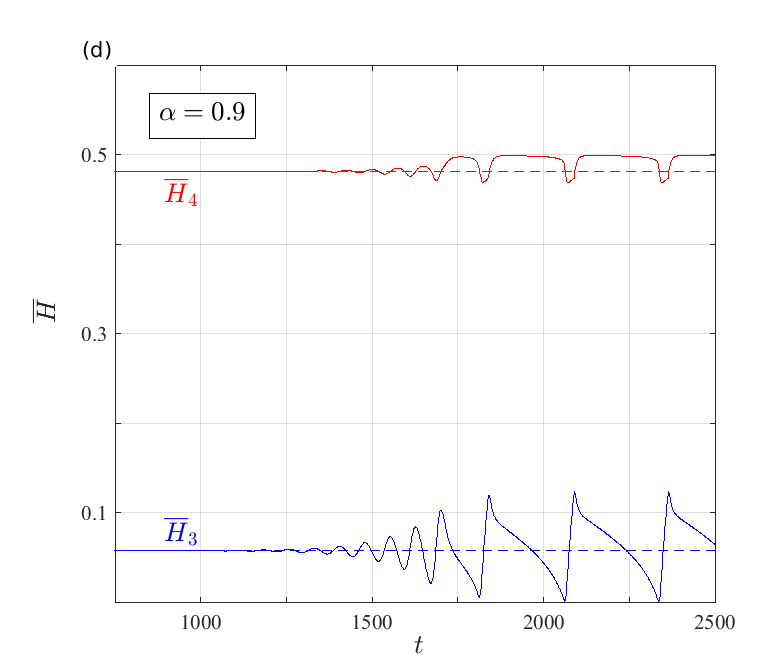}

\caption{Time evolution of $Q_3$, the fluid flux in vessel 3, and $\overline{H}_3$ and $\overline{H}_4$, the average haematocrit in vessels 3
and 4, respectively, showing how a nontrivial steady state (presented
by the dashed lines) undergoes a Hopf bifurcation to an oscillatory
solution (solid curves) for different values of the parameter $
\alpha$. In (a) $(\alpha,H_{0})=(0.5,0.5)$, while in
(b) $(\alpha,H_{0})=(0.9,0.5)$. Comparison of the solutions shows that as $\alpha$ approaches unity, the sensitivity of $\overline{H}_3$ and $\overline{H}_4$ to perturbations in $Q_3$ decreases.}

\label{Fig_feedback_q3_h_4}

\end{figure}

Consider, for example,
the feedback between the flow and haematocrit in vessel
3 (the redundant vessel) and vessel 4 when the flow in
vessel 3 is directed towards vessel 4 (oscillatory solutions for $\alpha\lesssim1$
and steady solutions for $\alpha \gtrsim 1$). The time evolution of the fluid flux
and haematocrit in this case are presented in Fig.~\ref{Fig_feedback_q3_h_4}
for $H_{0}=0.5$ and two values of $\alpha<1$. The following
positive feedback mechanism acts on the flux in
vessel 3: when $Q_{3}$ is increasing, more haematocrit enters vessel 3
(if the state of the system is located above the skimming threshold),
and less haematocrit enters vessel 4 (see the counter-phase
behaviour of $Q_{3}$ and $\overline{H}_{4}$ in Fig.~\ref{Fig_feedback_q3_h_4}).
In effect, the increase in $Q_{3}$ dilutes the haematocrit in vessel 4, while there is a time delay before the increase in the haematocrit entering vessel 3 propagates through the vessel and reaches vessel 4 (see the phase-lag between $Q_{3}$ and $\overline{H}_{3}$ in Fig.~\ref{Fig_feedback_q3_h_4}).
The decrease in haematocrit reduces the resistance in vessel 4, so that eventually more flow is redirected from the bottom to the top branch, through vessel 3. This half cycle is reversed when the
increase in the resistance of vessel 3 (due to the increase in its haematocrit)
becomes large enough to reduce $Q_{3}$ (see correspondence
between the maxima of $\overline{H}_{3}$ and the times at which $Q_{3}$
starts to decrease in Fig.~\ref{Fig_feedback_q3_h_4}). While similar
arguments can be applied when $\alpha>1$, changes in the haematocrit entering vessel 4 in response to perturbations in $Q_{3}$ are governed by the
ratio of the steady-state fluxes $Q_{3}^{(0)}/Q_{1}^{(0)}$ (see
Eq. (\ref{eq:h_4_non_trivial})). For $\alpha\lesssim1$, this ratio
is large enough to trigger a significant response; for larger values of $\alpha$, the ratio $Q_{3}^{(0)}/Q_{1}^{(0)}$ is smaller.
For example, in Fig.~\ref{Fig_feedback_q3_h_4}(a), $\alpha=0.5$ and
$Q_{3}^{(0)}/Q_{1}^{(0)}=0.853$, while in Fig.~\ref{Fig_feedback_q3_h_4}(b),
$\alpha=0.9$ and $Q_{3}^{(0)}/Q_{1}^{(0)}=0.045$. The reduction in
the ratio of fluid fluxes reduces the sensitivity of $\overline{H}_{4}$ to
perturbations in the redundant vessel (the amplitude of $\overline{H}_{4}$ reduces by 88\% in response to a 16\% reduction
in the amplitude of $Q_{3}$ between Fig.~\ref{Fig_feedback_q3_h_4}(a)
and (b)) and, thereby, suppresses the positive feedback mechanism.

In this work we chose the nominal diameter to be $D=20$ (20 $\mu m$ in dimensional units) as a representative diameter of physiological microcapillaries. Some quantitative differences should be expected when $D$ is changed. The critical value of $H_0$ at which oscillations emerge is dominated by the skimming threshold. When the nominal diameter $D$ is decreased, the dominant effect is a reduction in the value of $\alpha$ at which the critical $H_0$ is minimized ($\alpha \approx 0.55$ for $D=20$).
 This happens because, for a given haematocrit concentration, the change in resistance with the change in diameter ($\sim\mathrm{d}\left(\mu(H,D)/D^{4}\right)/\mathrm{d}D$)
increases as the diameter decreases. Therefore, the diameter difference between the two branches (i.e., the degree of structural asymmetry) decreases.
By contrast, the size of the stable region in a neighbourhood of $\alpha=1$ increases as $D$ decreases. In this case, the resistance of the redundant vessel (diameter $D$) increases and, therefore, a larger diameter ratio is needed to drive
flow through the redundant vessel.

While all results presented in this work are for equal inlet pressures and haematocrits, our calculations (not presented here for brevity) suggest that oscillatory solutions, having qualitatively similar dynamics to the oscillations presented here, exist when there is a relative inlet-pressure and haematocrit difference of a few percent. This shows that the oscillatory instability is not a unique feature of three-node networks with identical inlet conditions; rather it exists for a range of boundary conditions. The susceptibility of the three-node network to oscillatory dynamics for non-equal boundary conditions should help to realize these oscillations experimentally, as the identical inlet conditions are technically challenging 
to achieve in practice.

\subsection{Oscillations with a smooth haematocrit splitting function\label{subsec:Oscillations-_smooth}}

Motivated by the nonsmooth Hopf bifurcation to an oscillatory state encountered
when we use the haematocrit splitting model of \cite{Pri89},
it is natural to ask how smoothing the splitting
function will affect the stability of the system. Therefore, we use a third-order polynomial to smooth the
discontinuities in Eq.~(\ref{eq:Pries_model}) and consider an alternative splitting function of the form

\begin{equation}
F(\psi)=\begin{cases}
y_{L}(\psi), & \psi<\psi_{L}\\
\frac{e^{A}\left(\psi-\psi_{0}\right)^{B}}{e^{A}\left(\psi-\psi_{0}\right)^{B}+\left(1-\psi-\psi_{0}\right)^{B}} & \psi_{L}\leq\psi\leq\psi_{U}\\
y_{U}(\psi), & \psi>\psi_{U},
\end{cases}
\label{smooth_splitting_function}
\end{equation}

where $y_{L}(\psi)$ and $y_{U}(\psi)$ are third order polynomials and $\psi_{L}$ and $\psi_{U}$ represent points at which these polynomials intersect the original splitting function. A third-order polynomial
is the simplest functional form for which

\begin{equation}
F(\psi=0)=\frac{\partial F}{\partial\psi}|_{\psi=0}=\frac{\partial^{2}F}{\partial\psi^{2}}|_{\psi=0}=0;\label{eq:poly_conditions}
\end{equation}
these conditions are imposed in order to preserve the stability properties
of the trivial solution. We also require

\begin{equation}
\ \ \ \ \ \ \ \ \ \ \ F\ ,\ \frac{\partial F}{\partial\psi}\ \ \text{continuous on} \ \psi_{L}\ \text{and} \ \psi_{U}, \label{eq:poly_conditions_2}
\end{equation}
so that the governing equations are smooth throughout the parameter space.
The method used to construct smoothing polynomials that satisfy the
above conditions is described in Appendix C.

\begin{figure}[H]
\begin{centering}
\includegraphics[viewport=0bp 0bp 657bp 494.519bp,clip,scale=0.45]{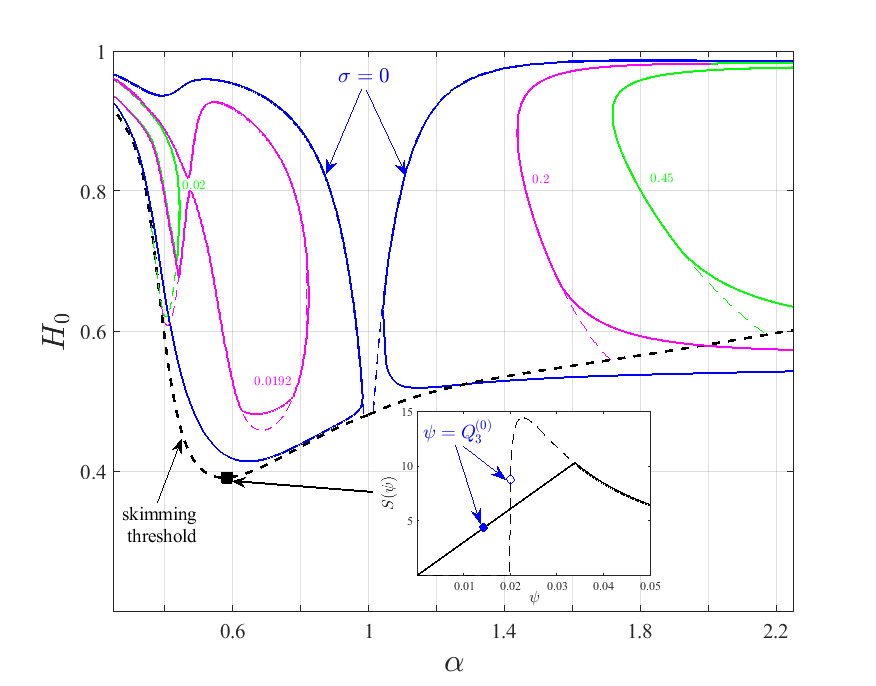}
\par\end{centering}
\caption{The effect of smoothing the \cite{Pri89} haematocrit
splitting function on the bifurcation to oscillatory solutions. The solid lines correspond to iso-growth-rate contours (the value of $\sigma$
is indicated) for the oscillatory solutions when the smooth splitting function (\ref{smooth_splitting_function}) is used,
while the dashed curves were produced using the original, nonsmooth
splitting function (Eq.~(\ref{eq:Pries_model}), as in Fig.~\ref{Fig_oscillatory_nonsmooth}). The iso-growth-rate
contours correspond to flow from the bottom to the top branch for $\alpha<1$
and vice versa for $\alpha>1$. The skimming threshold is plotted
for reference (thick dashed black line). The inset presents the function
$S(\psi)$ (see definition in Eq.~(\ref{eq:S})) at the location
marked by the black rectangle, $(\alpha,H_{0})=(0.58,0.391)$, for
the original, nonsmooth (dashed line) and smoothed (solid line) splitting functions.
The corresponding steady-state fluxes $Q_{3}^{(0)}$ obtained using
the nonsmooth (smooth) models at that parameter combination are indicated
with open (full) blue circles.}

\label{figure_stability_smooth}
\end{figure}

Figure~\ref{figure_stability_smooth} shows how the stability 
of the oscillatory solutions changes when the splitting function in Eq.~(\ref{eq:Pries_model}) is replaced by Eq. (\ref{smooth_splitting_function}). The results for the original, nonsmooth function are also included to facilitate comparison. Noticeably, the smooth
splitting function produces closed iso-$\sigma$ contours that differ from 
those obtained using the original, nonsmooth splitting function. Comparing the dashed and solid blue curves in Fig.~\ref{figure_stability_smooth}, smoothing the splitting function appears to increase stability for $\alpha\lesssim1.3$, while for larger values of $\alpha$ this trend is reversed. 
As expected, sufficiently far from the skimming
threshold, the smooth and nonsmooth haemaotcrit splitting rules yield identical stability results. 

Considering the perturbed equations of the nontrivial steady-state solutions, the haematocrit splitting rule appears only in Eq. (\ref{eq:h_3_non_trivial}), where it attains the following functional form 

\begin{equation}
S(\psi=Q_3^{(0)})=-\frac{1}{\psi}\left(\frac{F(\psi)}{\psi}-\frac{\partial F}{\partial\psi}\right)_{\psi=Q_3^{(0)}}.\label{eq:S}
\end{equation}

Therefore, the magnitude of $S(Q_{3}^{(0)})$ determines the size of $h_{3}/q_{3}$, the ratio
of the haematocrit to the fluid flux in the redundant vessel. As such, it plays a major role in the positive feedback mechanism (see discussion in Sec. \ref{subsec:Oscillatory-state}), associated with the onset of self-sustained oscillations.

The inset in Fig.~\ref{figure_stability_smooth} shows how differences between the smooth and nonsmooth splitting functions affect the behaviour at
the point $(\alpha,H_{0})=(0.58,0.391)$. This point is located just above the skimming threshold (unstable using the nonsmooth splitting rule), but below the curve on which $\sigma=0$ for the smooth splitting rule (stable). 
The function $S(\psi)$ shows how the choice of haematocrit splitting
rule affects the linear stability of nontrivial steady-state solutions and, therefore, how smoothing the splitting rule may change the system dynamics. The very large gradient of $S(\psi)$ in the nonsmooth model above the skimming-threshold also explains why the solution becomes unstable immediately above this critical value.
The different values of the steady-state fluxes
$Q_{3}^{(0)}$ obtained using the smooth and nonsmooth splitting functions at
$(\alpha,H_{0})=(0.58,0.391)$ are presented
in the inset. At these supercritical-skimming conditions ($F(Q_{3}^{(0)})>0$ for the nonsmooth model), the system with the nonsmooth splitting function produces a larger value of $S(Q_{3}^{(0)})$ than the system with the smooth splitting function, which results in an oscillatory instability of the former, while the latter is stable. 

From the experimental viewpoint, the functional form of
the haematocrit splitting rule for low flow rates (in a neighborhood of the skimming threshold) is challenging to measure because of the large noise-to-signal ratio that prevails at low flow rates. The stability results presented here for the cases of smooth and nonsmooth haematocrit splitting rules delineate the ``boundaries'' of possible behaviours for the given network. We postulate that other functional forms for smoothing
the model of \cite{Pri89} that satisfy
the conditions specified in (\ref{eq:poly_conditions}) and (\ref{eq:poly_conditions_2})
will yield bifurcation curves within these two ``bounds''.

\section{Conclusion\label{sec:Conclusion}}

In this paper we studied microcapillary blood-flow in a three-node network, exploring its multiple equilibria and the transition
to oscillations via dynamic simulations and stability analysis. While
multiple steady-state solutions and self-sustained oscillatory solutions
in microcapillary blood flows have been reported previously (see,
for example, \cite{KS15} and refs. cited therein), to our knowledge, the microstructural
characteristics that promote unsteady behaviour have not been identified. In this work we have demonstrated that specific structural abnormalities, in the form of redundant vessels which connect two flow paths with different resistances, are key to the emergence of oscillations. We 
have clarified the feedback mechanisms, arising due to the coupling of these structural features with the intrinsic nonlinearities of blood flow at the microscale (i.e., F\aa{}hr\ae{}us-Lindqvist effect and plasma skimming), which gives rise to oscillatory dynamics.

In our analysis, we defined a vessel as redundant if there is an equilibrium solution having zero-flow through that vessel. Such a ``trivial'' solution is typically unstable, with nontrivial steady-state solutions bifurcating (stable at the bifurcation point) for sufficiently large inlet haematocrit values. Remarkably, as we demonstrated using dynamic simulations, starting from the trivial
state, the system may evolve to either a different steady-state solution or an oscillatory solution. The paths leading to these long term solutions are sensitive to small changes in the inlet conditions, which
dictate the direction of flow in the redundant vessel. The sensitivity
of the system to small fluctuations in the boundary
conditions may lead to highly unstable behaviour if such a motif
is embedded in larger networks. Additionally, we postulate that the
maximum number of possible steady-state solutions should rise dramatically
as the number of redundant vessels in a network increases, because each redundant vessel may
support three solutions (no flow and/or flow in either direction).
The large number of equilibrium states, together with  sensitivity
to small fluctuations, may explain why highly irregular, almost chaotic flow is a characteristic feature of many vascular tumour networks \citep{Kim96, Bru07, Gil18}. 

To quantify the critical conditions for instability as the ratio of
branch diameters (representing the structural driving force) and
inlet haematocrit (representing the effect of local flow conditions) vary, we
performed stability analysis of the trivial solution. We found that
the transition from the trivial steady-state solution to oscillations occurs
in two steps -- the trivial state loses stability to a nontrivial
steady-state solution which, in turn, undergoes a Hopf-bifurcation.
By performing linear stability analysis of the nontrivial
steady-state solutions, we showed further that the combined effects of a redundant vessel and vessels that offer different resistances to flow (via different diameters in this work) is key to the emergence of self-induced oscillations. Also, we identified a feedback mechanism that facilitates the onset of oscillations; here, the diameter ratio between the two branches (affecting the flux in the redundant vessel) and the presence of haematocrit in the redundant vessel (allowing for time-delay in the system) are crucial ingredients for such positive feedback to occur. 
In future work, we aim to evaluate the effect of redundant vessels in larger vascular networks and to explore different motifs which may generate larger feedback loops and, thereby, larger scale oscillatory dynamics.
Such an investigation should consider
the coupled behaviour of multiple sources of oscillations, and how
their frequencies and amplitudes are modulated. Studying the haematocrit
oscillations in larger networks will ultimately enable us to evaluate their effect on tissue oxygenation, which is of considerable
importance in understanding the process of cycling hypoxia in tumours (as mentioned in Sec.~\ref{sec: introduction}).

Traditionally, studies of blood flow in large networks did not consider in detail what type of boundary conditions should be imposed because, in general, the appropriate choice of boundary conditions
for a microcapillary network is unknown. \cite{Fry12} showed that the choice of boundary conditions imposed on large-scale microcirculatory networks can significantly
influence the steady state flow rates. While most of the analysis in this paper was performed for constant pressure boundary conditions, we showed that changing to fixed-flux boundary conditions can destabilise the system, by reducing the critical inlet haematocrit at which the trivial solution becomes unstable.
Therefore, in future work, it would be of interest to examine how the stability of larger networks (where there are many more internal nodes than boundary nodes) is affected by changes in the type of boundary conditions imposed.

In this study we used a haematocrit splitting rule due to \cite{Pri89}; this model includes a threshold value of the daughter-to-parent
flux ratio, such that haematocrit only enters the daughter branch if the flow rate exceeds this critical value (the ``skimming threshold'').
The skimming threshold gives rise to a discontinuity in the splitting rule which, in turn, gives rise to nonsmooth stability diagrams. We used spline-smoothing
to eliminate the points of discontinuity in the model. In so doing, we obtained solutions which exhibited a smooth transition in parameter space between steady and oscillatory
states, while also converging to the results of the nonsmooth model
sufficiently far from the skimming threshold. In contrast to the steady-state solutions, the Hopf-bifurcation patterns are sensitive to small changes in the haematocrit splitting rules in the regime when the daughter-to-parent flux ratio is small (in the neighborhood of the skimming threshold). This sensitivity of the emergence of oscillatory dynamics to small changes in the haematocrit splitting rule used introduces significant challenges regarding how to measure and model the haematocrit splitting that occurs at such low flow rates.

\section*{Acknowledgments}
This work was supported by Cancer Research UK (CRUK)
Grant C47594/A29448, through the CRUK Oxford Center. G.W.A. acknowledges the support from the Engineering and Physical Research Council (Grant EP/L016044/1).

\section*{Data Availability}
The datasets generated during the current study are available from the corresponding author on reasonable request.

\bibliography{Literature_database}

\begin{thebibliography}{30}
\providecommand{\natexlab}[1]{#1}
\providecommand{\url}[1]{\texttt{#1}}
\expandafter\ifx\csname urlstyle\endcsname\relax
  \providecommand{\doi}[1]{doi: #1}\else
  \providecommand{\doi}{doi: \begingroup \urlstyle{rm}\Url}\fi

\bibitem[Bernabeu et~al.(2020)Bernabeu, K\"{o}ry, Grogan, Markelc, Beardo,
  d'Avezac, Enjalbert, Kaeppler, Daly, Hetherington, Kr\"{u}ger, Maini,
  Pitt-Francis, Muschel, Alarc\'{o}n, and Byrne]{Ber20}
M.~O. Bernabeu, J.~K\"{o}ry, J.~A. Grogan, B.~Markelc, A.~Beardo, M.~d'Avezac,
  R.~Enjalbert, J.~Kaeppler, N.~Daly, J.~Hetherington, T.~Kr\"{u}ger, P.~K.
  Maini, J.~M. Pitt-Francis, R.~J. Muschel, T.~Alarc\'{o}n, and H.~M. Byrne.
\newblock Abnormal morphology biases hematocrit distribution in tumor
  vasculature and contributes to heterogeneity in tissue oxygenation.
\newblock \emph{Proceedings of the National Academy of Sciences}, 117:\penalty0
  27811--27819, 2020.
\newblock \doi{10.1073/pnas.2007770117}.

\bibitem[Brurberg et~al.(2007)Brurberg, Benjaminsen, D\o{}rum, and
  Rofstad]{Bru07}
K.~G. Brurberg, I.~C. Benjaminsen, L.~M.~R. D\o{}rum, and E.~K. Rofstad.
\newblock Fluctuations in tumor blood perfusion assessed by dynamic
  contrast-enhanced mri.
\newblock \emph{Magnetic Resonance in Medicine}, 58:\penalty0 473--481, 2007.
\newblock \doi{10.1002/mrm.21367}.

\bibitem[Carr and Lacoin(2000)]{CL00}
R.~T. Carr and M.~Lacoin.
\newblock Nonlinear dynamics of microvascular blood flow.
\newblock \emph{Annals of Biomedical Engineering}, 28:\penalty0 641--652, 2000.
\newblock \doi{10.1114/1.1306346}.

\bibitem[Davis and Pozrikidis(2011)]{DP11}
J.~M. Davis and C.~Pozrikidis.
\newblock Numerical simulation of unsteady blood flow through capillary
  networks.
\newblock \emph{Bulletin of Mathematical Biology}, 73:\penalty0 1857--1880,
  2011.
\newblock \doi{10.1007/s11538-010-9595-3}.

\bibitem[Davis and Pozrikidis(2014{\natexlab{a}})]{DP14a}
J.~M. Davis and C.~Pozrikidis.
\newblock Self-sustained oscillations in blood flow through a honeycomb
  capillary network.
\newblock \emph{Bulletin of Mathematical Biology}, 76:\penalty0 2217--2237,
  2014{\natexlab{a}}.
\newblock \doi{10.1007/s11538-014-0002-3}.

\bibitem[Davis and Pozrikidis(2014{\natexlab{b}})]{DP14b}
J.~M. Davis and C.~Pozrikidis.
\newblock On the linear stability of blood flow through model capillary
  networks.
\newblock \emph{Bulletin of Mathematical Biology}, 76:\penalty0 2985--3015,
  2014{\natexlab{b}}.
\newblock \doi{10.1007/s11538-014-0041-9}.

\bibitem[F\aa{}hr\ae{}us(1929)]{Far29}
R.~F\aa{}hr\ae{}us.
\newblock The suspension stability of the blood.
\newblock \emph{Physiological Reviews}, 9:\penalty0 241--274, 1929.
\newblock \doi{10.1152/physrev.1929.9.2.241}.

\bibitem[F\aa{}hr\ae{}us and Lindqvist(1931)]{FL31}
R.~F\aa{}hr\ae{}us and T.~Lindqvist.
\newblock The viscosity of blood in narrow capillary tubes.
\newblock \emph{American Journal of Physiology}, 96:\penalty0 562--568, 1931.

\bibitem[Fenton et~al.(1985)Fenton, Carr, and Cokelet]{FCC85}
B.~M. Fenton, R.~T. Carr, and G.~R. Cokelet.
\newblock Nonuniform red cell distribution in 20 to 100 $\mu$m bifurcations.
\newblock \emph{Microvascular Research}, 29:\penalty0 103--126, 1985.
\newblock \doi{10.1016/0026-2862(85)90010-X}.

\bibitem[Forouzan et~al.(2012)Forouzan, Yang, M.Sosa, Burns, and
  Shevkoplyas]{Fro12}
O.~Forouzan, X.~Yang, J.~M.Sosa, J.~M. Burns, and S.~S. Shevkoplyas.
\newblock Spontaneous oscillations of capillary blood flow in artificial
  microvascular networks.
\newblock \emph{Microvascular Research}, 84:\penalty0 123--132, 2012.
\newblock \doi{10.1016/j.mvr.2012.06.006}.

\bibitem[Fry et~al.(2012)Fry, Lee, Smith, and Secomb]{Fry12}
B.~C. Fry, J.~Lee, N.~P. Smith, and T.~W. Secomb.
\newblock Estimation of blood flow rates in large mcrovascular networks.
\newblock \emph{Microcirculation}, 19:\penalty0 530--538, 2012.
\newblock \doi{10.1111/j.1549-8719.2012.00184.x}.

\bibitem[Gardner et~al.(2010)Gardner, Li, Small, Geddes, and Carr]{Gar10}
D.~Gardner, Y.~Li, B.~Small, J.~B. Geddes, and R.~T. Carr.
\newblock Multiple equilibrium states in a micro-vascular network.
\newblock \emph{Mathematical Biosciences}, 227:\penalty0 117--124, 2010.
\newblock \doi{10.1016/j.mbs.2010.07.001}.

\bibitem[Geddes et~al.(2007)Geddes, Carr, Karst, and Wu]{Ged07}
J.~B. Geddes, R.~T. Carr, N.~J. Karst, and F.~Wu.
\newblock The onset of oscillations in microvascular blood flow.
\newblock \emph{SIAM Journal on Applied Dynamical Systems}, 6:\penalty0
  694--727, 2007.
\newblock \doi{10.1137/060670699}.

\bibitem[Geddes et~al.(2010)Geddes, Carr, Wu, Lao, and Maher]{Ged10}
J.~B. Geddes, R.~T. Carr, F.~Wu, Y.~Lao, and M.~Maher.
\newblock Blood flow in microvascular networks: A study in nonlinear biology.
\newblock \emph{Chaos}, 20:\penalty0 045123, 2010.
\newblock \doi{10.1063/1.3530122}.

\bibitem[Gillies et~al.(2018)Gillies, Brown, Anderson, and Gatenby]{Gil18}
R.~J. Gillies, J.~S. Brown, A.~R.~A. Anderson, and R.~A. Gatenby.
\newblock Eco-evolutionary causes and consequences of temporal changes in
  intratumoural blood flow.
\newblock \emph{Nature Reviews Cancer}, 18:\penalty0 576--585, 2018.
\newblock \doi{10.1038/s41568-018-0030-7}.

\bibitem[Gray et~al.(1953)Gray, Conger, Ebert, Hornsey, and Scott]{Gra53}
L.~H. Gray, A.~D. Conger, M.~Ebert, S.~Hornsey, and O.~C.~A. Scott.
\newblock The concentration of oxygen dissolved in tissues at the time of
  irradiation as a factor in radiotherapy.
\newblock \emph{The Britishe Journal of Radiology}, 26:\penalty0 638--648,
  1953.
\newblock \doi{10.1259/0007-1285-26-312-638}.

\bibitem[Harrison and Blackwell(2004)]{HB04}
L.~Harrison and K.~Blackwell.
\newblock Hypoxia and anemia: factors in decreased sensitivity to radiation
  therapy and chemotherapy?
\newblock \emph{The Oncologist}, 9:\penalty0 31--40, 2004.

\bibitem[H\"{o}ckel et~al.(1996)H\"{o}ckel, Schlenger, Aral, Mitze,
  Sch\"{a}ffer, and Vaupel]{Hoc96}
M.~H\"{o}ckel, K.~Schlenger, B.~Aral, M.~Mitze, U.~Sch\"{a}ffer, and P.~Vaupel.
\newblock Association between tumor hypoxia and malignant progression in
  advanced cancer of the uterine cervix.
\newblock \emph{Cancer Research}, 56:\penalty0 4509--4515, 1996.

\bibitem[Horsman et~al.(2012)Horsman, Mortensen, Petersen, Busk, and
  Overgaard]{Hor12}
M.~R. Horsman, L.~S. Mortensen, J.~B. Petersen, M.~Busk, and J.~Overgaard.
\newblock Imaging hypoxia to improve radiotherapy outcome.
\newblock \emph{Nature Reviews Clinical Oncology}, 9:\penalty0 674--687, 2012.
\newblock \doi{10.1038/nrclinonc.2012.171}.

\bibitem[Jain(2005)]{Jain05}
R.~K. Jain.
\newblock Normalization of tumor vasculature: An emerging concept in
  antiangiogenic therapy.
\newblock \emph{Science}, 307:\penalty0 58--62, 2005.
\newblock \doi{10.1126/science.1104819}.

\bibitem[Karst et~al.(2015)Karst, Storey, and Geddes]{KS15}
N.~J. Karst, B.~Storey, and J.~B. Geddes.
\newblock Oscillations and multiple equilibria in microvascular blood flow.
\newblock \emph{Bulletin of Mathematical Biology}, 77:\penalty0 1377--1400,
  2015.
\newblock \doi{10.1007/s11538-015-0089-1}.

\bibitem[Kiani et~al.(1994)Kiani, Pries, Hsu, Sarelius, and Cokelet]{Kia94}
M.~F. Kiani, A.~R. Pries, L.~L. Hsu, I.~H. Sarelius, and G.~R. Cokelet.
\newblock Fluctuations in microvascular blood flow parameters caused by
  hemodynamic mechanisms.
\newblock \emph{American Journal of Physiology-Heart and Circulatory
  Physiology}, 266:\penalty0 H1822--H1828, 1994.
\newblock \doi{10.1152/ajpheart.1994.266.5.H1822}.

\bibitem[Kimura et~al.(1996)Kimura, Braun, Ong, Hsu, Secomb, Papahadjopoulos,
  Hong, and Dewhirst]{Kim96}
H.~Kimura, R.~D. Braun, E.~T. Ong, R.~Hsu, T.~W. Secomb, D.~Papahadjopoulos,
  K.~Hong, and M.~W. Dewhirst.
\newblock Fluctuations in red cell flux in tumor microvessels can lead to
  transient hypoxia and reoxygenation in tumor parenchyma.
\newblock \emph{Cancer Research}, 56:\penalty0 5522--5528, 1996.

\bibitem[Klitzman and Johnson(1982)]{KJ82}
B.~Klitzman and P.~C. Johnson.
\newblock Capillary network geometry and red cell distribution in hamster
  cremaster muscle.
\newblock \emph{American Journal of Physiology}, 242:\penalty0 H211--H219,
  1982.
\newblock \doi{10.1152/ajpheart.1982.242.2.H211}.

\bibitem[Krogh(1921)]{Kro21}
A.~Krogh.
\newblock Studies on the physiology of capillaries: {I}{I}. the reactions to
  local stimuli of the blood-vessels in the skin and web of the frog.
\newblock \emph{The Journal of Physiology}, 55:\penalty0 412--422, 1921.
\newblock \doi{10.1113/jphysiol.1921.sp001985}.

\bibitem[Michiels et~al.(2016)Michiels, Tellie, and Feron]{CTF16}
C.~Michiels, C.~Tellie, and O.~Feron.
\newblock Cycling hypoxia: A key feature of the tumor microenvironment.
\newblock \emph{Biochimica et Biophysica Acta (BBA)-Reviews on Cancer},
  1866:\penalty0 76--86, 2016.
\newblock \doi{10.1016/j.bbcan.2016.06.004}.

\bibitem[Pries et~al.(1989)Pries, Ley, Claassen, and Gaehtgens]{Pri89}
A.~R. Pries, K.~Ley, M.~Claassen, and P.~Gaehtgens.
\newblock Red cell distribution at microvascular bifurcations.
\newblock \emph{Microvascular Research}, 38:\penalty0 81--101, 1989.
\newblock \doi{10.1016/0026-2862(89)90018-6}.

\bibitem[Pries et~al.(1992)Pries, Fritzsche, Ley, and Gaehtgens]{Pri92}
A.~R. Pries, A.~Fritzsche, K.~Ley, and P.~Gaehtgens.
\newblock Redistribution of red blood cell flow in microcirculatory networks by
  hemodilution.
\newblock \emph{Circulation Research}, 70:\penalty0 1113--1121, 1992.
\newblock \doi{10.1161/01.RES.70.6.1113}.

\bibitem[Pries et~al.(1994)Pries, Secomb, Gessner, Sperandlo, Gross, and
  Gaehtgens]{Pea94}
A.~R. Pries, T.~W. Secomb, T.~Gessner, M.~B. Sperandlo, J.~F. Gross, and
  P.~Gaehtgens.
\newblock Resistance to blood flow in microvessels in vivo.
\newblock \emph{Circulation Research}, 75:\penalty0 904--915, 1994.
\newblock \doi{10.1161/01.RES.75.5.904}.

\bibitem[Storey et~al.(2015)Storey, Hellen, Karst, and Geddes]{Sto15}
B.~D. Storey, D.~V. Hellen, N.~J. Karst, and J.~B. Geddes.
\newblock Observations of spontaneous oscillations in simple two-fluid
  networks.
\newblock \emph{Physical Review E}, 91:\penalty0 023004, 2015.
\newblock \doi{10.1103/PhysRevE.91.023004}.

\end{thebibliography}
\newpage
\section*{Appendix}

\subsection*{A -- Linear stability equations for steady-state solutions with
$Q_{3}$ directed towards the upper branch (Case \RomanNumeralCaps{1})\label{subsec:App_A}}

Having specified the haematocrit in vessels 3 and 4 in Eqs.~(\ref{eq:h_3_non_trivial})
and (\ref{eq:h_4_non_trivial}), we write the $O(\epsilon)$ haematocrit mass balance
(\ref{eq:junction_mass_balance}) at nodes 2-3-5 and 4-5-6 as follows:

\begin{equation}
h_{5}=H_{0}\frac{\left(F(Q_{3}^{(0)})-Q_{3}^{(0)}\right)}{\left(1-Q_{3}^{(0)}\right)^{2}}\left[q_{2}-\frac{q_{3}}{Q_{3}^{(0)}}\right]-\frac{Q_{3}^{(0)}}{1-Q_{3}^{(0)}}h_{3}\label{eq:App_A_first}
\end{equation}

and

\begin{equation}
\begin{gathered}\begin{aligned}h_{6}= & \frac{H_{0}}{\left(1+Q_{1}^{(0)}\right)^{2}}\left[\frac{Q_{1}^{(0)}+F(Q_{3}^{(0)})}{Q_{1}^{(0)}+Q_{3}^{(0)}}-\frac{1-F(Q_{3}^{(0)})}{1-Q_{3}^{(0)}}\right]\\
 & \times\left[\left(1-Q_{3}^{(0)}\right)q_{1}-\left(Q_{1}^{(0)}+Q_{3}^{(0)}\right)q_{2}+\left(1+Q_{1}^{(0)}\right)q_{3}\right]\\
 & +\frac{Q_{1}^{(0)}+Q_{3}^{(0)}}{1+Q_{1}^{(0)}}h_{4}\exp\left(-\frac{\alpha^{2}}{Q_{1}^{(0)}+Q_{3}^{(0)}}\lambda\right)\\
 & +\frac{1-Q_{3}^{(0)}}{1+Q_{1}^{(0)}}h_{5}\exp\left(-\frac{\lambda}{1-Q_{3}^{(0)}}\right).
\end{aligned}
\end{gathered}
\end{equation}

The pressure drop along the loop formed between the three internal nodes yields Eq.~(\ref{eq:triangle_steady_Q3_negative}) at zero order, while at $O(\epsilon)$

\begin{equation}
\begin{aligned}\frac{1}{\mu(H_{0},D)}\left[\frac{\mu(H_{4}^{(0)},\alpha D)}{\alpha^{4}}q_{1}-\mu(H_{5}^{(0)},D)q_{2}+\left(\mu(H_{3}^{(0)},D)+\frac{\mu(H_{4}^{(0)},\alpha D)}{\alpha^{4}}+\mu(H_{5}^{(0)},D)\right)q_{3}\right]\\
+Q_{3}^{(0)}r_{3}+(Q_{1}^{(0)}+Q_{3}^{(0)})r_{4}-(1-Q_{3}^{(0)})r_{5} & =0.
\end{aligned}
\end{equation}

Since the pressure differences imposed between the inlet and outlet nodes are fixed, we can assume that the $O(\epsilon)$ perturbations to these pressure differences are equal to zero. Therefore, the $O(\epsilon)$ terms of Eq.~(\ref{dp_1_and_2}) yield

\begin{equation}
\frac{\mu(H_{0}^{(0)},\alpha D)}{\mu(H_{0},D)\alpha^{4}}q_{1}-q_{2}-\frac{\mu(H_{3}^{(0)},D)}{\mu(H_{0},D)}q_{3}-Q_{3}^{(0)}r_{3}=0,
\end{equation}

and

\begin{equation}
\begin{aligned}\left[1+\frac{1}{\mu(H_{0},D)\alpha^{4}}\left(\mu(H_{0},\alpha D)+\mu(H_{4}^{(0)},\alpha D)\right)\right]q_{1}+q_{2}+\frac{\mu(H_{4}^{(0)},\alpha D)}{\mu(H_{0},D)\alpha^{4}}q_{3}\\
+\left(Q_{1}^{(0)}+Q_{3}^{(0)}\right)r_{4}+\left(1+Q_{1}^{(0)}\right)r_{6} & =0.
\end{aligned}
\label{eq:App_A_last}
\end{equation}

Equations~(\ref{eq:App_A_first})-(\ref{eq:App_A_last}), together
with Eqs.~(\ref{eq:h_3_non_trivial}) and (\ref{eq:h_4_non_trivial}),
and the relation between vessel haematocrit and resistance in (\ref{eq:resistance_series}),
form the eigenvalue problem in (\ref{eq:A_v}) for the case when flow in the redundant vessel is directed towards the top branch (Case \RomanNumeralCaps{1}).

\subsection*{B -- Linear stability equations for steady-state solutions with
$Q_{3}$ directed towards the bottom branch (Case \RomanNumeralCaps{2})}

We write the $O(\epsilon)$ perturbations to the steady-state
solution given by Eqs.~(\ref{eq:trinagle_steady_Q3_positive})-(\ref{eq:H_steady_Q3_positive}). At $O(\epsilon)$, Eq.~(\ref{eq:H_steady_Q3_positive}) yields

\begin{equation}
h_{3}=H_{0}\left(\frac{F^{*}(Q_{3/1}^{(0)})}{Q_{3/1}^{(0)}}-\frac{\partial F^{*}}{\partial\psi}|_{Q_{3/1}^{(0)}}\right)\left[q_{1}-\frac{q_{3}}{Q_{3/1}^{(0)}}\right],
\label{eq:App_B_first}
\end{equation}

\begin{equation}
h_{4}=H_{0}\frac{F^{*}(Q_{3/1}^{(0)})-Q_{3/1}^{(0)}}{Q_{1}^{(0)}\left(1-Q_{3/1}^{(0)}\right)^{2}}\left[q_{1}-\frac{q_{3}}{Q_{3/1}^{(0)}}\right]-\frac{Q_{3/1}^{(0)}}{1-Q_{3/1}^{(0)}}h_{3},
\end{equation}

and

\begin{equation}
h_{5}=H_{0}Q_{1}^{(0)}\frac{Q_{3/1}^{(0)}-F^{*}(Q_{3/1}^{(0)})}{\left(1+Q_{3}^{(0)}\right)^{2}}\left[q_{2}-\frac{q_{3}}{Q_{3}^{(0)}}\right]+\frac{Q_{3}^{(0)}}{1+Q_{3}^{(0)}}h_{3}\exp\left(-\frac{\lambda}{Q_{3}^{(0)}}\right).
\end{equation}

The $O(\epsilon)$ of the haematocrit mass balance at node 4-5-6 yields

\begin{equation}
\begin{gathered}\begin{aligned}h_{6}= & \frac{H_{4}^{(0)}-H_{5}^{(0)}}{\left(1+Q_{1}^{(0)}\right)^{2}}\left[\left(1+Q_{3}^{(0)}\right)q_{1}+\left(Q_{1}^{(0)}-Q_{3}^{(0)}\right)q_{2}-\left(1+Q_{1}^{(0)}\right)q_{3}\right]\\
 & +\frac{Q_{1}^{(0)}-Q_{3}^{(0)}}{1+Q_{1}^{(0)}}h_{4}\exp\left(-\frac{\alpha^{2}}{Q_{1}^{(0)}-Q_{3}^{(0)}}\lambda\right)\\
 & +\frac{1+Q_{3}^{(0)}}{1+Q_{1}^{(0)}}h_{5}\exp\left(-\frac{\lambda}{1+Q_{3}^{(0)}}\right).
\end{aligned}
\end{gathered}
\end{equation}

The pressure drop along the loop formed between the three internal nodes yields at $O(\epsilon)$,

\begin{equation}
\begin{aligned}\frac{1}{\mu(H_{0},D)}\left[-\frac{\mu(H_{4}^{(0)},\alpha D)}{\alpha^{4}}q_{1}+\mu(H_{5}^{(0)},D)q_{2}+\left(\mu(H_{3}^{(0)},D)+\frac{\mu(H_{4}^{(0)},\alpha D)}{\alpha^{4}}+\mu(H_{5}^{(0)},D)\right)q_{3}\right]\\
+Q_{3}^{(0)}r_{3}-(Q_{1}^{(0)}-Q_{3}^{(0)})r_{4}+(1+Q_{3}^{(0)})r_{5} & =0.
\end{aligned}
\end{equation}

Since the pressure differences imposed between the inlet and outlet nodes are fixed, we can assume that the $O(\epsilon)$ perturbations to these pressure differences are equal to zero. Therefore, the $O(\epsilon)$ terms of Eq.~(\ref{dp_1_and_2}) yield

\begin{equation}
\frac{\mu(H_{0}^{(0)},\alpha D)}{\mu(H_{0},D)\alpha^{4}}q_{1}-q_{2}+\frac{\mu(H_{3}^{(0)},D)}{\mu(H_{0},D)}q_{3}+Q_{3}^{(0)}r_{3}=0,
\end{equation}

and

\begin{equation}
\begin{aligned}\left[1+\frac{1}{\mu(H_{0},D)\alpha^{4}}\left(\mu(H_{0},\alpha D)+\mu(H_{4}^{(0)},\alpha D)\right)\right]q_{1}+q_{2}-\frac{\mu(H_{4}^{(0)},\alpha D)}{\mu(H_{0},D)\alpha^{4}}q_{3}\\
+\left(Q_{1}^{(0)}-Q_{3}^{(0)}\right)r_{4}+\left(1+Q_{1}^{(0)}\right)r_{6} & =0.
\end{aligned}
\label{eq:App_B_last}
\end{equation}

Equations~(\ref{eq:App_B_first})-(\ref{eq:App_B_last}), together
with the relation between vessel haematocrit and resistance in (\ref{eq:resistance_series}),
form the eigenvalue problem in (\ref{eq:A_v}) for the case when flow in the redundant vessel is directed towards the bottom branch (Case \RomanNumeralCaps{2}).

\subsection*{C -- Calculation of the smoothing third-order polynomial}

The smoothed haematocrit splitting function takes the form introduced in Eq.~(\ref{smooth_splitting_function}), in which the polynomials that satisfy the conditions specified in (\ref{eq:poly_conditions})
and (\ref{eq:poly_conditions_2}) read

\begin{equation}
\begin{gathered}y_{L}(\psi)=a_{L}\psi^{3}\;\;\;\text{for}\;\;\;\psi\leq\psi_{L}\;\;\;\text{and}\\
y_{U}(\psi)=1+a_{U}\left(\psi-1\right)^{3}\;\;\;\text{for}\;\;\;\psi\geq\psi_{U},
\end{gathered}
\end{equation}

where $\psi_{L}$ and $\psi_{U}$ are the points of intersection of $y_{L}(\psi)$ and $y_{U}(\psi)$, respectively, with the original splitting functions (Eq.~(\ref{smooth_splitting_function})).
These points of intersection are found by numerically solving the
nonlinear equations

\begin{equation}
\frac{\psi_{L}}{3}-\frac{F(\psi_{L})}{\partial F/\partial\psi|_{\psi_{L}}}=0\;\;\;\text{and}\;\;\;\frac{\psi_{U}-1}{3}-\frac{F(\psi_{U})-1}{\partial F/\partial\psi|_{\psi_{U}}}=0,
\end{equation}

which allow us to calculate the polynomial coefficients as 
\[
a_{L}=\frac{F(\psi_{L})}{\psi_{L}^{3}}\ \ \ \text{and}\ \ \ a_{U}=\frac{F(\psi_{U})}{\left(\psi_{U}-1\right)^{3}}.
\]

\end{document}